\documentclass[11pt,a4paper]{article}

\usepackage[a4paper,margin=1in]{geometry}
\usepackage[T1]{fontenc}
\usepackage[utf8]{inputenc}
\usepackage{lmodern}
\usepackage{microtype}
\usepackage{graphicx}
\usepackage{amsmath,amssymb}
\usepackage{booktabs}
\usepackage{enumitem}
\usepackage{xcolor}
\usepackage{float}
\usepackage{hyperref}
\hypersetup{
  colorlinks=true,
  linkcolor=blue!60!black,
  citecolor=blue!60!black,
  urlcolor=blue!60!black
}
\usepackage{array}
\newcolumntype{C}[1]{>{\centering\arraybackslash}p{#1}}
\newcolumntype{L}[1]{>{\raggedright\arraybackslash}p{#1}}

\usepackage{tikz}

\usepackage{amsmath}

\usepackage[ruled,vlined,linesnumbered]{algorithm2e}

\newtheorem{theorem}{Theorem}[section]

\newtheorem{definition}[theorem]{Definition}
\newtheorem{example}[theorem]{Example}

\newtheorem{remark}[theorem]{Remark}

\title{A Topological Framework for Atmospheric River Interaction Using Framed Braids}

\author{Ioannis Diamantis\\
\small Department of Data Analytics and Digitalisation, Maastricht University\\
\small Maastricht, The Netherlands\\
\small \texttt{i.diamantis@maastrichtuniversity.nl}
}

\date{} % add a date if you like

\begin{document}
\maketitle

\begin{abstract}
Atmospheric Rivers (ARs) are filamentary moisture pathways responsible for a large fraction of extreme precipitation and often occur as interacting filament bundles within the same synoptic regime. Existing diagnostics typically analyze
ARs in isolation, despite the frequent coexistence and interaction of multiple filaments. We introduce a topological framework for AR analysis based on framed braids and framed braidoids, which encodes both the geometric interaction of AR
centroids and the internal evolution of moisture transport.

In this approach, AR filaments are represented as strands whose time-ordered crossings form braid words, while moisture-based framing captures internal intensification or weakening along each filament. Applying this framework to reanalysis-derived Atmospheric River track data, we construct braid and framed braid representations over sliding time windows and analyze a strongly interacting multi-filament AR episode in the North Pacific. The results show that braid-based indicators capture structural reorganizations and moisture intensification episodes that are not apparent from centroid geometry or IVT magnitude alone, offering a complementary structural perspective on atmospheric moisture transport.
\end{abstract}

\section{Introduction}

Atmospheric Rivers (ARs) are long, narrow filamentary structures that transport a large fraction of poleward water vapor in the midlatitudes
\cite{ZhuNewell1998, Rutz2019, RalphDettinger2012}. They account for a substantial portion of extreme precipitation and flooding across many regions of the world, with major implications for hydrology, water-resource management, and climate variability \cite{GuanWaliser2015, Gorodetskaya2020, Lavers2011}. ARs are commonly analyzed using Eulerian fields such as Integrated Vapor Transport (IVT) or via Lagrangian centroid trajectories and Lagrangian Coherent Structure diagnostics \cite{Haller2015}. However, such approaches typically treat AR filaments as isolated objects. In reality, multiple ARs often coexist within the same synoptic regime and may interact, merge, bifurcate, or undergo rapid structural reconfiguration driven by jet streaks, baroclinic development, and Rossby-wave dynamics \cite{Payne2020, Deflorio2018, Baggett2016}.

A growing body of work has begun to characterize AR geometry and evolution using skeletonization techniques, filament-tracking algorithms, Eulerian metrics, and dynamical systems tools
\cite{Rutz2019, EirasBarca2023, Cao2020}. While these methods capture intensity, shape, and persistence, they do not explicitly encode the \emph{interaction topology} of multiple AR filaments—namely, the time-ordered reorganization of their relative positions, crossings, and structural exchanges as part of a coherent moisture-transport network \cite{Dacre2019, GuanWaliser2015}. Understanding such interaction-driven structure is important because it often accompanies rapid AR intensification, fragmentation, or reorientation.

In this work we introduce a topological framework for AR analysis based on braid theory, framed braids, and framed braidoids. A braid encodes the time-ordered crossings of multiple trajectories; a \emph{framed braid} enriches each strand with an integer-valued structure capturing internal variation; and a \emph{braidoid} generalizes these constructions by allowing strands to appear or disappear. These tools have proven powerful in dynamical systems, fluid mixing, and mathematical physics
\cite{Boyland2000, Thiffeault2010, FinnThiffeault2006, Kauffman1991}, but have not previously been used to model the structural interaction of Atmospheric River filaments or their internal moisture evolution.

We represent each AR filament as a moisture-bearing strand whose centroid evolves in time and whose internal structure is quantified using an IVT-based moisture integral \cite{GuanWaliser2015}. When several ARs evolve simultaneously, their pathways may undergo geometric crossings while their moisture content strengthens, weakens, or shears. Encoding these processes using algebraic generators, i.e. crossings $\sigma_i$ and framings $\tau_i$, yields a compact symbolic description of AR evolution that is robust to smooth geometric deformation yet sensitive to physically meaningful reorganization. Using reanalysis-based AR track data, we extract braid representations over sliding time windows, extend them to framed braids via moisture diagnostics, and illustrate their interpretive value through detailed case studies of multi-filament AR interactions.

\smallbreak 
The remainder of the paper is organized as follows.
Section~2 reviews the geometric and physical structure of Atmospheric Rivers and introduces the moisture-based quantities used in the topological model. Section~3 presents the necessary background on braid theory, framed braids,
and framed braidoids, together with their physical interpretation in the AR context. In Section~4 we describe the extraction of oriented braid representations from reanalysis-derived AR centroid tracks, while Section~5
extends this construction to framed braids using moisture diagnostics. Section~6 presents case studies of interacting multi-filament AR events and discusses the resulting topological indicators. Finally, limitations and future directions are discussed in Section~7.

%%%%%%%%%%%%%%%%%%%%%%%%%%%%%%%%%%%%%%%%%%%%%%%%%%%%%

\section{Atmospheric Rivers: Geometry, Dynamics, and Moisture Structure}

Atmospheric Rivers (ARs) are long, filamentary corridors of concentrated water vapor transport embedded within the midlatitude westerlies \cite{ZhuNewell1998, Rutz2019, GuanWaliser2015}. Despite occupying a small fraction of the subtropical and extratropical atmosphere, they dominate poleward moisture transport and play a leading role in heavy precipitation, flooding, and hydrological extremes \cite{Dettinger2013, RalphDettinger2012, Payne2020}. To build a topological model of AR evolution using braid theory and its framed extensions, it is essential to first establish a geometric and structural description that captures the intrinsic physical characteristics of these filaments.

\subsection{Geometric description of ARs}

An Atmospheric River is typically defined as a coherent corridor of enhanced moisture transport exceeding roughly 2,000 km in length and 200–500 km in width, identified using intensity thresholds in Integrated Vapor Transport (IVT) or Integrated Water Vapor (IWV) \cite{RutzSteenburgh2014, GuanWaliser2015}. The central axis of this corridor, commonly referred to as the centroid line or core trajectory, delineates the path of maximum moisture flux and provides a one-dimensional representation of the filament around which the AR band is organized \cite{Neiman2008, Maclennan2025}. This centroid serves as the natural mathematical object for representing AR evolution in the braid formulation introduced later.

\begin{figure}
    \centering
    \includegraphics[width=0.6\linewidth]{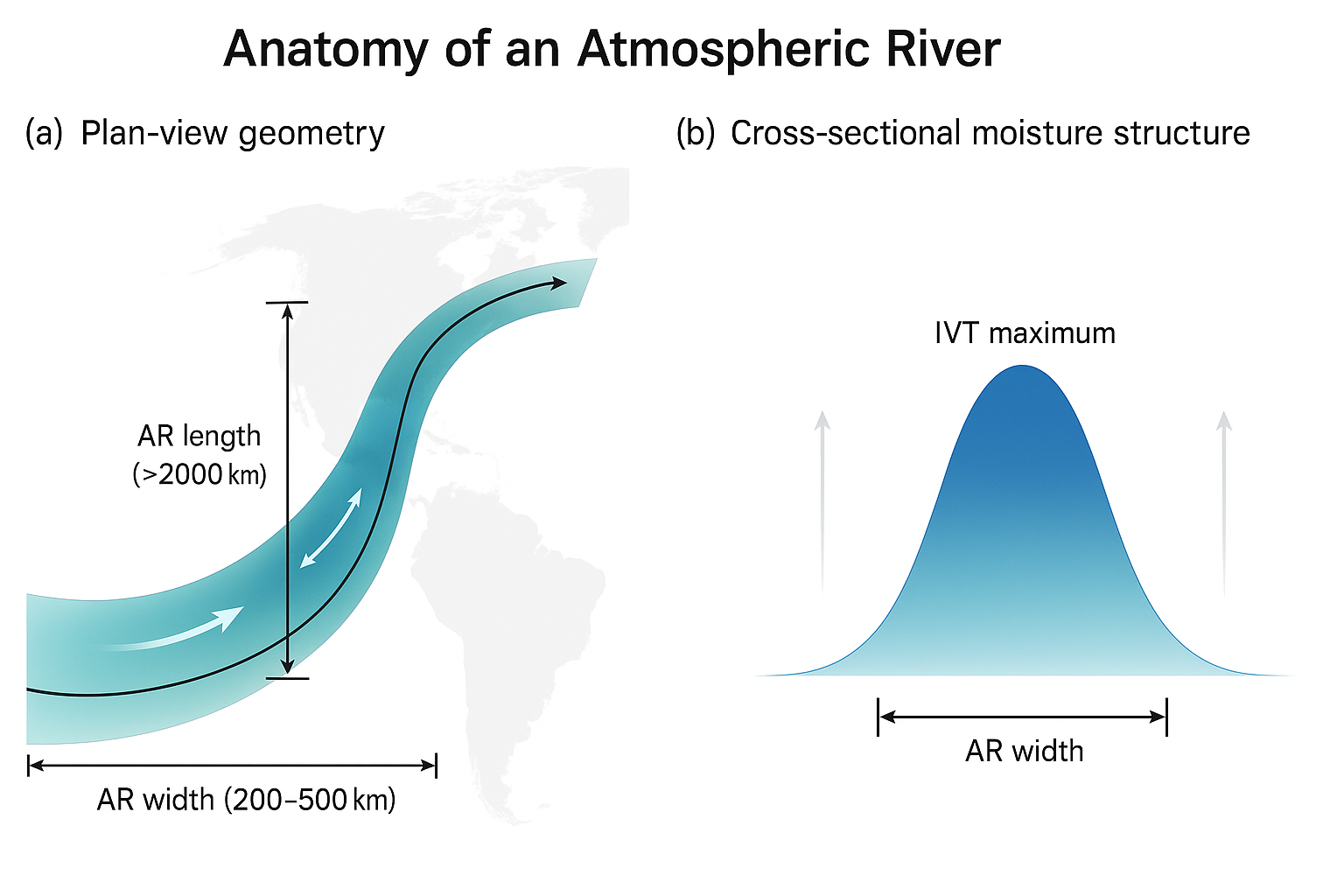}
    \caption{Anatomy of an Atmospheric River. (a) Plan-view geometry showing the centroid and IVT-defined band. (b) Cross-sectional IVT profile illustrating the moisture-rich core and AR width.}
    \label{fig:ar}
\end{figure}

Although centroid evolution is generally smooth on synoptic scales, the surrounding dynamical environment often introduces localized structural changes. Interactions with extratropical cyclones, transient jet streaks, atmospheric blocking, and planetary-wave disturbances can induce sharp curvature, sudden inflections, localized thinning or thickening, and occasional short-lived bifurcations of the filament backbone \cite{Han2021, PayneMagnusdottir2014, Gorodetskaya2020}. Such geometric deviations play a central role in the braid-based description developed below. The relative ordering of multiple AR centroids in the zonal direction determines the generators of the braid representation, and changes in this ordering correspond to topological crossings whose sequence encodes the geometric complexity of the AR system. We emphasize that such crossings reflect changes in longitudinal ordering of centroids rather than physical intersection of moisture filaments, and serve as a symbolic encoding of geometric reorganization rather than literal contact.

\subsection{Cross-sectional structure and the band model}

While the centroid provides a one-dimensional characterization of an Atmospheric River, its physical impacts arise from the two-dimensional cross-sectional moisture distribution surrounding this backbone. At each point along the centroid, the perpendicular cross-section of the IVT field typically exhibits an elongated and anisotropic structure with a dominant moisture maximum \cite{Maclennan2025, Li2024}. The region where IVT exceeds a defined threshold therefore forms a continuous band enveloping the centroid \cite{GuanWaliser2019, Rutz2019}. This band representation captures key geometric and structural attributes that govern AR-induced precipitation and hydrological influence.

The band possesses a measurable width, reflecting the geometric thickness of the corridor through which moisture is efficiently transported \cite{RutzSteenburgh2014}. More importantly for our purposes, the cross-sectional distribution of IVT encodes a degree of internal shear in the moisture transport field and moisture loading that varies along the filament \cite{Maclennan2025}. This quantity can drift, intensify or localize in response to dynamical forcing, and provides the physical interpretation for the framing of a strand in the topological model developed below. The outer boundaries of the band form coherent transport edges separating the AR from surrounding lower-moisture regions, and their evolution influences how the filament interacts with ambient flow fields \cite{GuanWaliser2015}.

The internal structure of ARs is far from rigid. Moisture maxima can shift laterally, split into multiple cores or exhibit rotation driven by upper-level dynamics, jet-induced shear or potential-vorticity gradients \cite{Gorodetskaya2020, Dacre2019}. These deformations not only modify the band geometry but also introduce local twisting and shear that can be naturally represented as variations in the framing of a thickened strand. In this sense, the AR band carries both geometric information (its centroid and width) and structural information (its internal moisture state), and framed braid theory provides a direct algebraic mechanism to encapsulate both components.

\subsection{Dynamical evolution and interactions}

Atmospheric Rivers rarely evolve as isolated features. Their structure, intensity, and trajectory continuously respond to the surrounding synoptic environment. When an AR interacts with an extratropical cyclone, the resulting warm conveyor belt flow can stretch and intensify the moisture corridor, whereas frontal wave development may fracture the filament into two or more branches \cite{PayneMagnusdottir2014, Baggett2016}. Anticyclonic blocking can impose abrupt directional changes as the filament is deflected along the block periphery, occasionally generating highly curved or stalled pathways \cite{Gorodetskaya2020}. Moisture plumes originating from subtropical regions may join existing ARs, forming multi-core structures with enhanced transport \cite{RalphDettinger2012}. Breaking Rossby waves and jet streak variations can reorganize AR geometry more rapidly, leading to transient folds, bifurcations, or overlap between filaments \cite{Han2021}.

\begin{figure}
    \centering
    \includegraphics[width=0.6\linewidth]{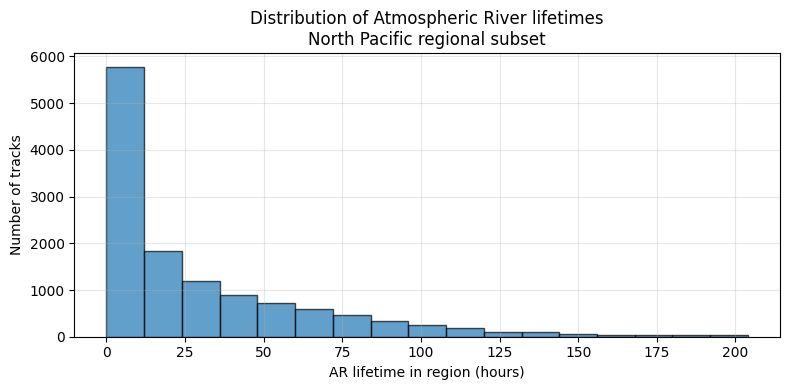}
    \caption{Distribution of AR lifetimes within the North Pacific domain. Longer-lived filaments are more likely to undergo interactions and structural reorganization.}
    \label{fig:distr}
\end{figure}

These dynamical influences give rise to distinct structural transitions. When two filaments approach each other under differential advection, their one-dimensional centroids may, in principle, overtake one another, or exchange longitudinal order. During strong intensification, internal shear or jet-induced rotation can conceptually create localized twisting within the moisture band. When environmental forcing weakens a portion of a filament, a previously continuous corridor may, in some cases, split into separate segments, or a secondary branch may emerge. Conversely, two nearby filaments may merge into a single intensified moisture corridor when guided by coherent large-scale flow. Finally, individual short-lived filaments may appear and dissipate entirely within a larger multi-filament environment as local moisture anomalies form and decay.

These transformations are geometrically coherent yet temporally discrete, altering the connectivity or relative positioning of the moisture pathways. This combination of smooth evolution punctuated by configuration changes is what makes ARs amenable to a topological representation. The topological framework introduced in the next section formalizes how these reorganizations can be encoded and analyzed, allowing both geometric interaction and internal structure to be represented within a unified mathematical model.

Consequently, AR evolution can be viewed as a time-ordered sequence of framed braid operations induced by synoptic-scale atmospheric dynamics. The braid-based perspective does not alter the physical drivers of AR behavior, but provides a symbolic and structurally robust representation of how multiple moisture filaments reorganize, interact, and transform over time.

\begin{figure}
    \centering
    \includegraphics[width=0.6\linewidth]{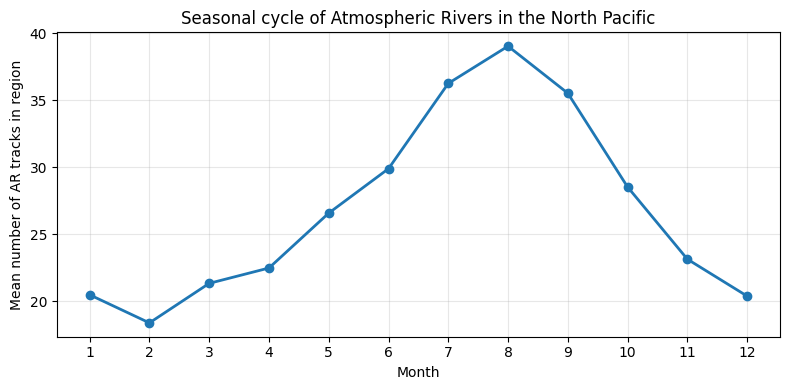}
    \caption{Seasonal cycle of AR track occurrences within the selected regional domain, showing the mean number of detected AR filaments per month.}
    \label{fig:Season}
\end{figure}

\subsection{Moisture accumulation and the role of framing}

Although ARs are commonly visualized as one-dimensional centroids, their physical impacts arise from the evolving two-dimensional moisture field surrounding the core trajectory. Local intensification of IVT within the cross-section increases the condensation potential of the filament and is strongly linked to downstream precipitation extremes \cite{Ralph2004, Maclennan2025, GuanWaliser2015}. To encode
these internal structural changes along the AR in a strand-attached manner, we associate to each point on the centroid a measure of moisture load derived from the IVT field in the perpendicular direction.

\begin{definition}[Moisture load along an AR filament]
Let $s$ denote arc length along an Atmospheric River centroid, and let $\Sigma(s)$ be a cross-sectional slice of the AR band perpendicular to the centroid at $s$. We define the \emph{moisture load} at $s$ by
\begin{equation}\label{fram}
M(s) \;\propto\; \int_{\Sigma(s)} \mathrm{IVT}(x,y,s)\,\mathrm{d}A ,
\end{equation}
where $\mathrm{IVT}$ denotes the standard scalar magnitude of integrated vapor transport.
\end{definition}

\begin{remark}\rm
Quantities based on cross-sectionally integrated IVT have appeared previously in the atmospheric literature as diagnostics of moisture transport strength. Here, however, $M(s)$ is attached to the evolving AR centroid and is introduced specifically as a strand-level structural variable whose discretized variation will be interpreted as a framing in a topological braid model.
\end{remark}

Large values of $M(s)$ correspond to compact, moisture-loaded segments with strong internal gradients, whereas reductions reflect weakening, broadening, or partial detachment of the filament \cite{Neiman2008, EirasBarca2023}. In practice, we discretize $\Sigma(s)$ on the native grid and normalize the resulting integral to ensure comparability across ARs and across datasets with different spatial resolutions.

This scalar description of internal structure is consistent with widely used atmospheric diagnostics, such as IVT maxima and gradients \cite{GuanWaliser2015}, IWV concentrations \cite{Li2024}, meridional moisture flux anomalies
\cite{RutzSteenburgh2014}, and PV-driven deformation and rotation of the plume \cite{Hoskins1985}, but differs in that it provides a trajectory-based, rather than purely Eulerian, representation. By associating $M(s)$ with the evolving
centroid, we obtain a one-dimensional sequence that captures how internal moisture organization strengthens, reorganizes, or dissipates as the AR interacts with its environment.

\begin{remark}\rm
In the topological representation developed later in this paper, the moisture-based quantity $M(s)$ plays the role of a \emph{framing} variable along each strand. In classical framed braid theory, the framing records how a thickened strand twists around its core curve \cite{KoSmo1}. Here, this abstract notion is interpreted physically: variations in internal moisture content and cross-sectional shear act
as a proxy for this twist. The values of $M(s)$ are normalized so that an order-one change corresponds to a physically meaningful variation in moisture transport,
ensuring robustness under changes in grid resolution and comparability across AR events.
\end{remark}

\subsection{Why ARs admit a topological model}

Although Atmospheric Rivers are dynamical fluid structures, their evolution exhibits properties that are naturally described from a topological perspective. AR filaments maintain a coherent identity over thousands of kilometers and over multi-day timescales \cite{Rutz2019}. For extended periods they undergo smooth and continuous evolution, yet their most significant structural changes, including bifurcation, merging, sudden curvature shifts and filament decay, occur as discrete events imposed by the surrounding synoptic environment \cite{Han2021}. When multiple ARs coexist, their trajectories may interweave or exchange longitudinal order in a manner that reflects structured geometric interaction rather than random variability \cite{Lavers2011, Deflorio2018}. At the same time, their physical characteristics, including length, width, orientation and internal moisture organization, are geometry-dependent properties that evolve coherently along the continuous backbone of the filament \cite{RutzSteenburgh2014, Maclennan2025}. Taken together, these features represent precisely the type of behavior that topological models are designed to capture: continuous deformation interrupted by discrete reorganizations that alter connectivity or relative positioning.

A topological representation therefore offers a viewpoint that complements, rather than replaces, dynamical formulations and statistical descriptions of ARs. By focusing on structural evolution (how for example filaments interact, split, merge and intensify) it isolates the organizational backbone of AR behavior in a mathematically tractable form. The framework developed in the next sections formalizes this perspective by representing AR centroids and their internal moisture structure as strands whose evolution over time can be encoded symbolically. This provides a compact way to characterize structural transitions in the moisture-transport network and lays the foundation for new diagnostics of AR organization and interaction.

\section{Framed Braids and Braidoids: Mathematical Background}

This section introduces the topological structures used in our representation of Atmospheric Rivers (ARs). Classical braid theory \cite{Artin, Birman1975} provides the foundation for encoding the time-ordered crossings of multiple AR centroids. To account for the finite width and internal structure of ARs, each strand is thickened into a band, leading to the theory of framed braids \cite{KoSmo1}. However, the number of AR filaments present in a region varies over time as they appear, dissipate, split, or merge. These processes require the more general framework of \emph{framed braidoids}, which permits strands with free endpoints and possible splitting and merging behavior. In the subsections below, we outline these constructions rigorously while maintaining a physical interpretation aligned with AR dynamics.

\subsection{Classical braids}\label{subsec:classical_braids}

A classical braid on $n$ strands is a collection of $n$ disjoint smooth curves embedded in $\mathbb{R}^2 \times [0,1]$ such that each curve monotonically increases in the vertical direction. The strands do not intersect in three-dimensional space, although their projections to $\mathbb{R}^2$ may have transversal crossings.

\begin{figure}
\centering
\includegraphics[width=0.3\linewidth]{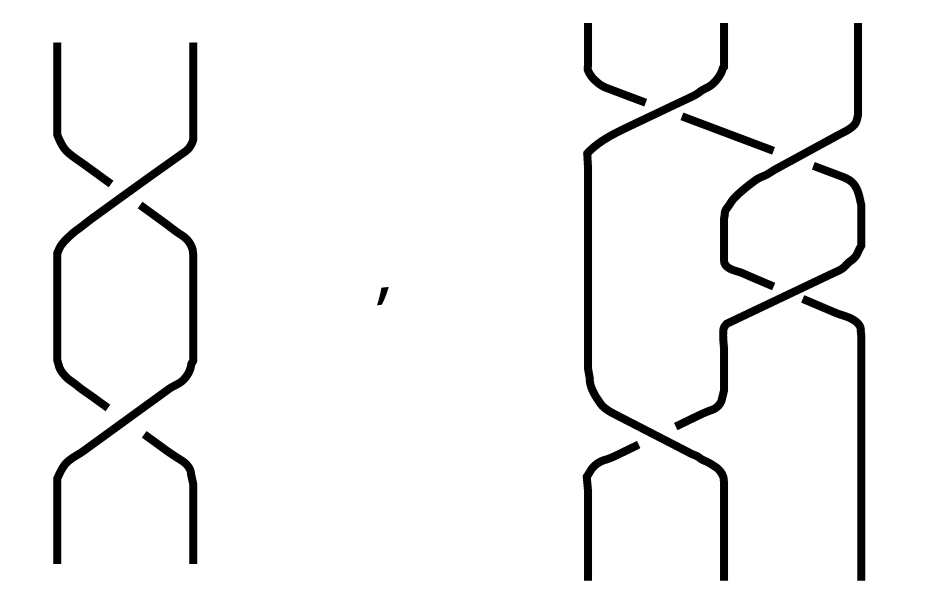}
\caption{Examples of a 2--strand and a 3--strand braid. Time increases upward.}
\label{fig:basic_braids}
\end{figure}

Two braids are considered equivalent if one can be smoothly deformed into the other within the surrounding space $\mathbb{R}^2 \times [0,1]$ without cutting or intersecting strands, without reversing the direction of time, and while keeping the endpoints fixed on the boundary planes. Such a deformation is called an \emph{ambient isotopy}.
 \footnote{Informally, one braid can be ``wiggled'' into the other without cutting or reconnecting strands, reversing time, or altering the attachment points at $t=0$ and $t=1$.
}.

\begin{remark}\rm 
An important operation on a braid is 
\emph{closure}: connecting each strand's top endpoint with its corresponding bottom endpoint by non-intersecting arcs in $\mathbb{R}^3$. This produces a link in $S^3$, and by Alexander's theorem every link can be represented as the closure of a braid \cite{Alexander1923}.
\end{remark}

\begin{figure}
\centering
\includegraphics[width=0.4\linewidth]{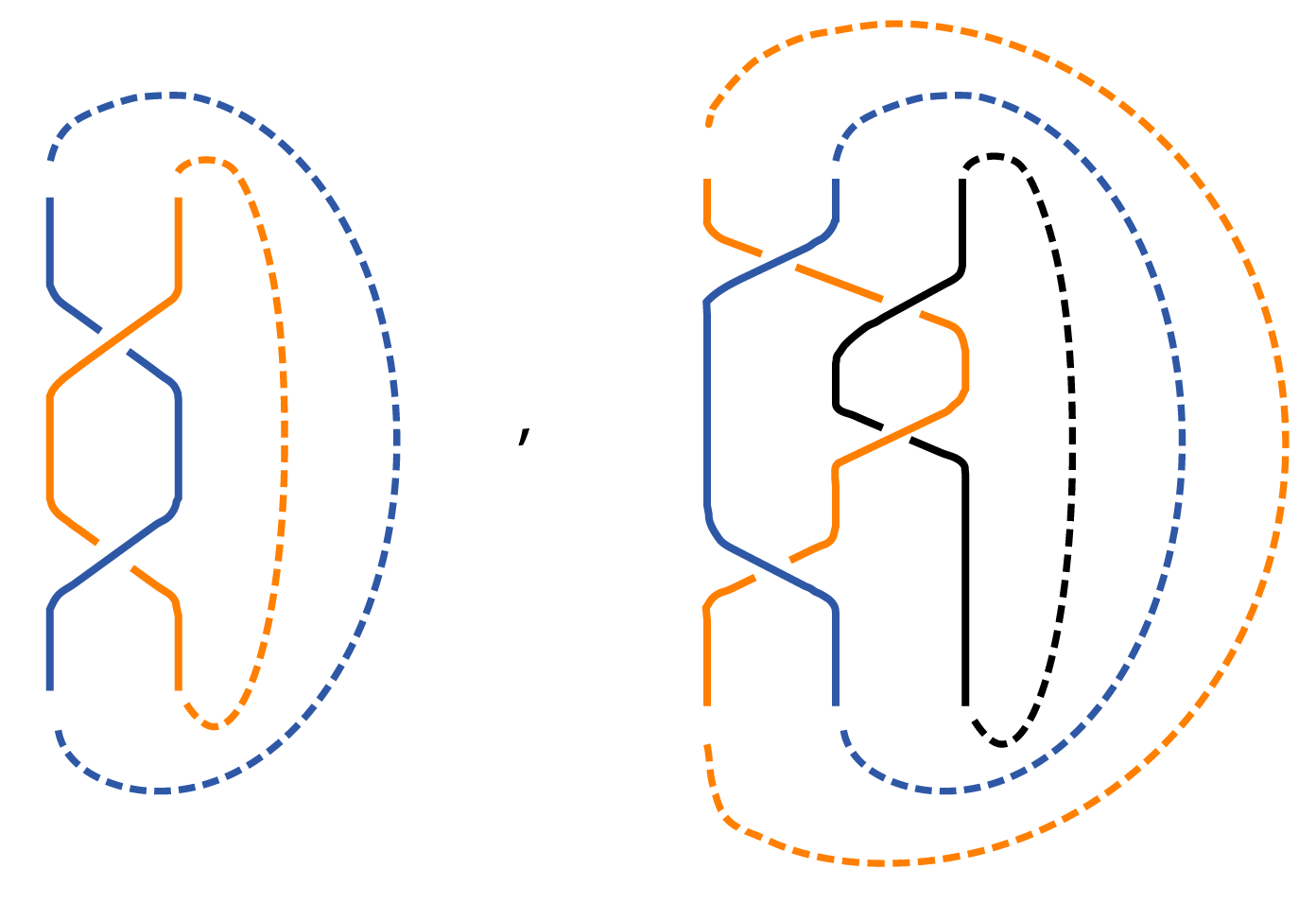}
\caption{Closure of a 2-strand and a 3-strand braid producing two links.}
\label{fig:closure}
\end{figure}

The set of all $n$--strand braids admits a natural composition operation given by vertical concatenation (stacking): the bottom endpoints of one braid are identified with the top endpoints of the next, and the resulting embedding is again a braid. Smooth deformations (ambient isotopies) that preserve the disjointness of strands, the monotonic time direction, and the endpoints do not change the essential structure of a braid. When braids are considered up to such deformations, the resulting equivalence classes form a group, called the \emph{braid group} $B_n$.

A classical presentation of $B_n$ is given by the Artin generators
\[
\sigma_1, \ldots, \sigma_{n-1},
\]
where $\sigma_i$ represents strand $i$ passing transversely over 
strand $i{+}1$, and $\sigma_i^{-1}$ represents the corresponding 
under-crossing. These satisfy the defining relations
\[
\sigma_i \sigma_j = \sigma_j \sigma_i 
\quad \text{if } |i-j|>1,
\qquad
\sigma_i \sigma_{i+1}\sigma_i
=
\sigma_{i+1}\sigma_i\sigma_{i+1}.
\]
Every element of $B_n$ can be represented by a \emph{braid word} in the $\sigma_i^{\pm 1}$, with different words representing the same element related by the relations above.

\begin{figure}[H]
\centering
\includegraphics[width=0.5\linewidth]{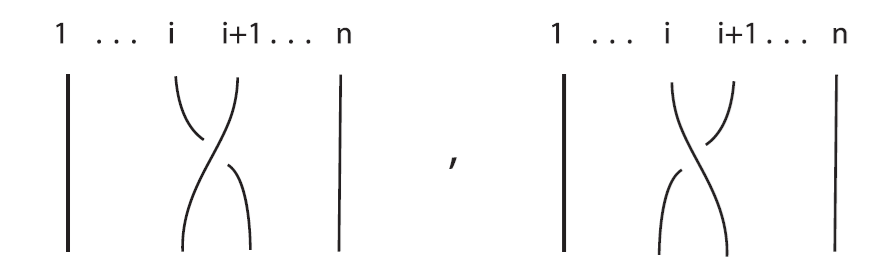}
\caption{Braid group generators $\sigma_i$ (right) and $\sigma_i^{-1}$ (left).}
\label{fig:generators}
\end{figure}

The group structure formally encodes the time-ordered exchange of lateral positions between strands. In our atmospheric application, these strands will represent centroid trajectories of AR filaments, and the resulting braid words symbolically capture the geometric interactions among them.

The algebraic relations in $B_n$ correspond to the classical Reidemeister moves of type 2 and 3, which allow for the local deformation of strand projections without altering the underlying topology.

\begin{figure}[H]
\centering
\includegraphics[width=0.7\linewidth]{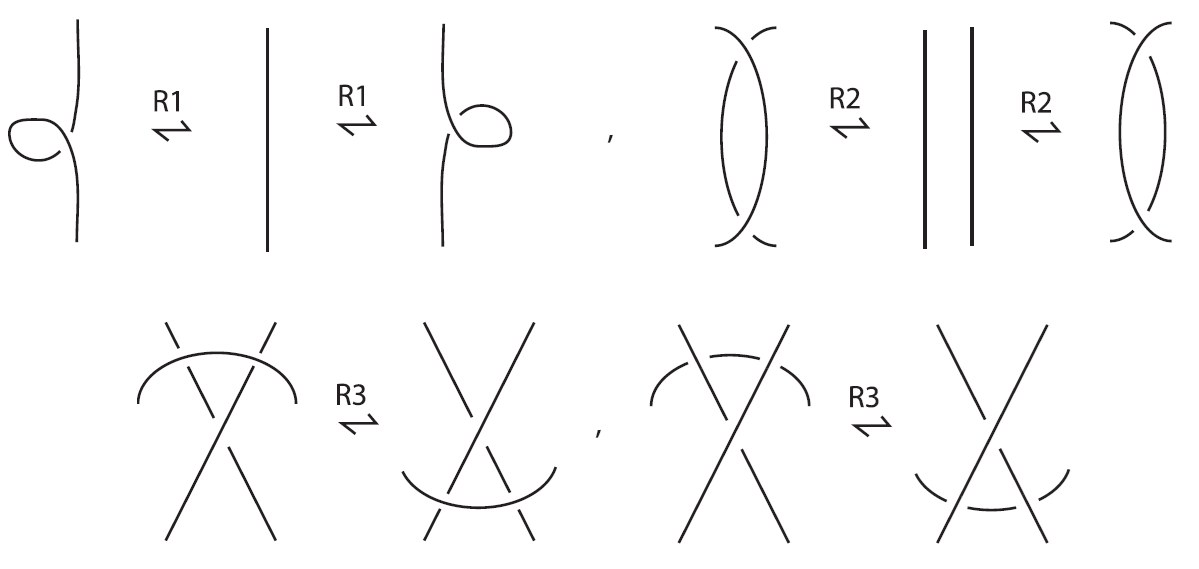}
\caption{The Reidemeister moves.}
\label{fig:R_moves}
\end{figure}

\begin{figure}[H]
\begin{center}
\includegraphics[width=5.5in]{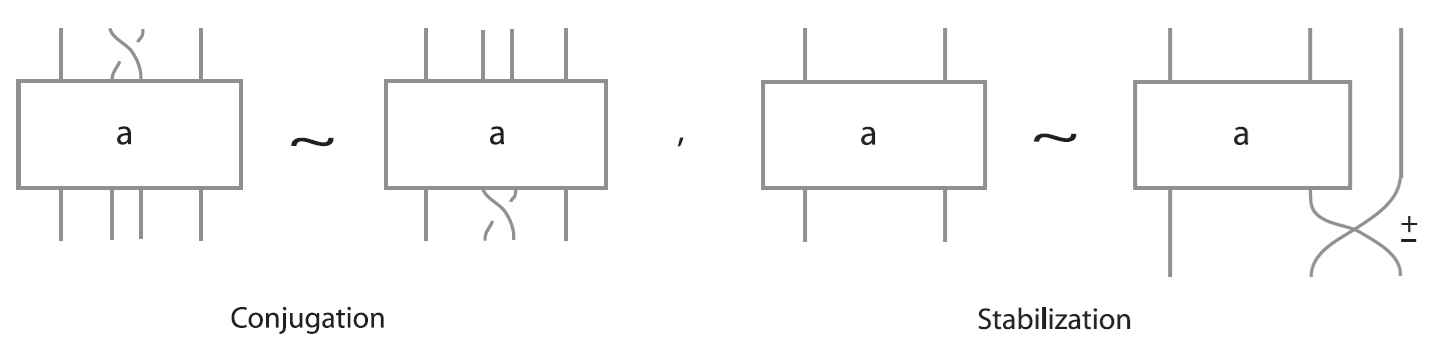}
\end{center}
\caption{Braid equivalence.}
\label{beq}
\end{figure}

Finally, it is worth mentioning that equivalence classes of braids are in one-to-one correspondence with the isotopy classes of oriented links. More precisely:

\begin{theorem}[Markov]\label{Mark}
The closures of two braids are isotopic if and only if one braid can be taken to another by finite sequence of the following moves (for an illustration see Figure~\ref{beq}):
\[
\begin{array}{lllll}
{\rm Conjugation:} & a & \sim & bab^{-1}, & a, b \in B_n, \\
&&&\\
{\rm Stabilization:} & a  & \sim & a\sigma_n^{-1}, & a \in B_n. \\
\end{array}
\]
\end{theorem}

\subsection{Framed braids}

Classical braids represent one-dimensional strand trajectories. To model objects of finite width and internal structure, each strand may be replaced by a disjoint embedded band (or ribbon) in $\mathbb{R}^2 \times [0,1]$, producing a \emph{framed braid}. The framing records the twisting of each band about its core curve and is measured as an integer. While the framing is integer-valued in the topological setting, its physical interpretation will be defined via a discretization of a continuous moisture-based quantity in Section~\ref{subsec:physicalframe}.

\begin{figure}[H]
\centering
\includegraphics[width=0.35\linewidth]{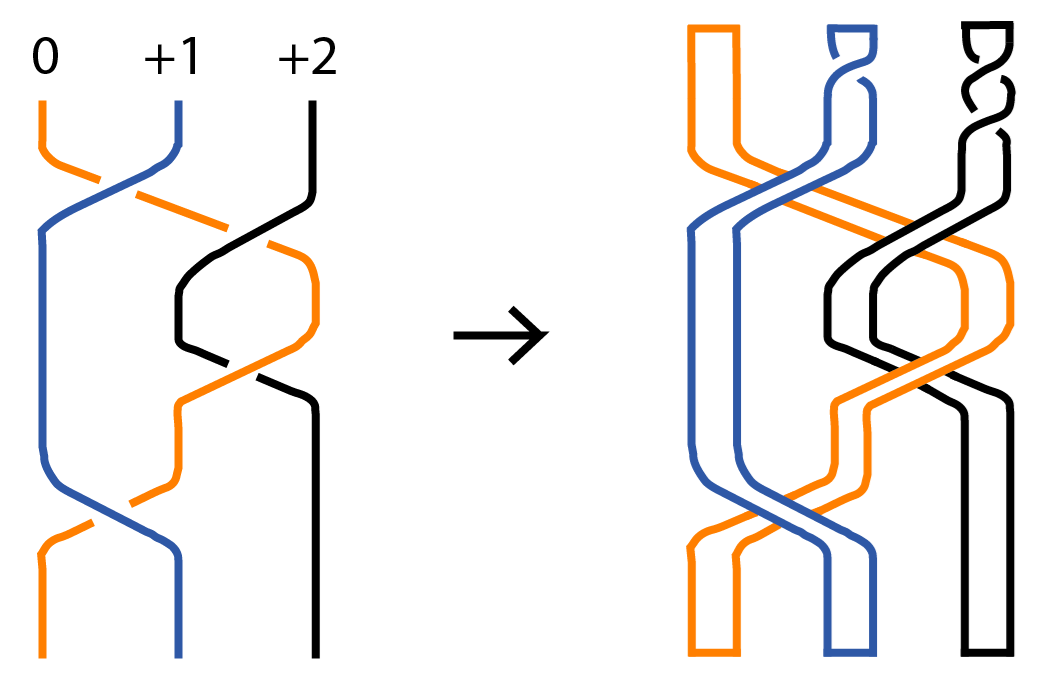}
\caption{Examples of integer framings ($0$, $+1$, $+2$ twists) applied to the three strands of a braid.}
\label{fig:framed_braid}
\end{figure}
 
The algebraic theory of framed braids is due to Ko \& Smolinsky \cite{KoSmo1}, who showed that the framed braid group on $n$ strands can be expressed as a semidirect product
\[
FB_n \;\cong\; \mathbb{Z}^n \rtimes B_n,
\]
where the $B_n$ component encodes strand interactions and the $\mathbb{Z}^n$ component tracks the integer twisting of each band under strand exchanges. For generalizations to surface braids see \cite{BellingeriGervais2012}.

Let $\sigma_i$ denote the standard Artin generators of the braid group $B_n$, and let $\tau_i$ denote a full twist of the $i$--th band about its core. Then $FB_n$ admits the presentation:
\[
FB_n
=
\left\langle
\sigma_1,\dots,\sigma_{n-1},\,
\tau_1,\dots,\tau_n
\;\middle|\;
\begin{aligned}
%&\text{(Artin relations)} \\[2pt]
&\sigma_i\sigma_j = \sigma_j\sigma_i \quad (|i-j|>1), \\[-2pt]
&\sigma_i\sigma_{i+1}\sigma_i = \sigma_{i+1}\sigma_i\sigma_{i+1}, \\[6pt]
%&\text{(Framing relations)} \\[2pt]
&\tau_i\tau_j = \tau_j\tau_i, \\[4pt]
&\sigma_i \tau_i = \tau_{i+1}\sigma_i, \\[-2pt]
&\sigma_i \tau_{i+1} = \tau_i \sigma_i, \\[-2pt]
&\sigma_i \tau_j = \tau_j \sigma_i 
\quad (j \neq i,i{+}1),
\end{aligned}
\right\rangle.
\]

\begin{figure}[H]
\centering
\includegraphics[width=0.5\linewidth]{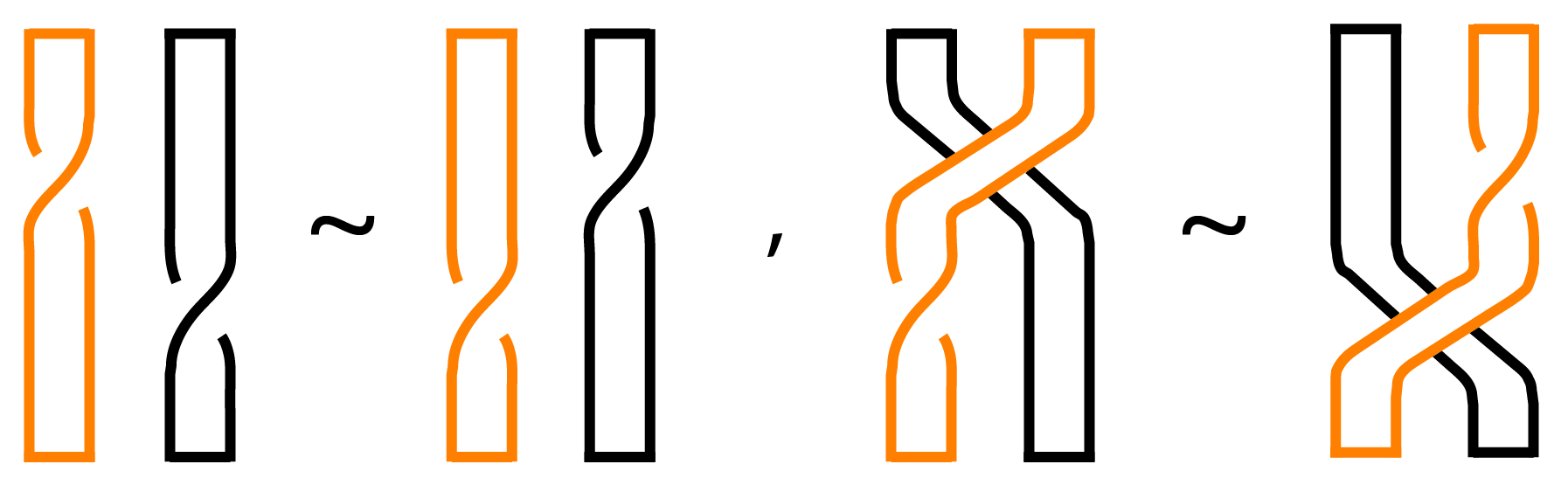}
\caption{
Left: the relations between twists:
$\tau_i \tau_j\, =\, \tau_j \tau_i$, for $i\neq j$.
Right: the relations between twists and crossings:
$\tau_i\sigma_i\, =\, \sigma_i \tau_{i+1}$.
}
\label{fig:tau_generator}
\end{figure}

The generators $\sigma_i$ encode the time-ordered exchange of lateral positions between strands, while the $\tau_i$ measure the internal twisting of each band. In the atmospheric interpretation developed later, these twist values correspond to physically meaningful variations in internal moisture structure along AR filaments.

\subsection{Physical interpretation of framing}\label{subsec:physicalframe}

As mentioned above, in framed braid theory, the integer framing associated with a strand measures the number of full rotations of an embedded band about its core curve. This topological twist quantifies structural information beyond the geometry of the strand. For Atmospheric Rivers, we reinterpret the same concept in physical terms: 

\bigbreak 

\noindent {\it The ``band'' represents the 
two-dimensional moisture corridor surrounding the AR centroid, and its internal rotation or compaction corresponds to spatially integrated changes 
in vapor transport.}

\bigbreak 

Let $s$ denote arc length along the centroid and let $\Sigma(s)$ be the perpendicular cross-section through the moisture corridor at that location. We associate to each point on the centroid the scalar $M(s)$ defined in 
Equation~\eqref{fram}, which reflects the total IVT transported through the cross-section. Regions of enhanced $M(s)$ indicate a moist, compact filament with strong transverse gradients, while decreases signify spreading, weakening, or partial detrainment of moisture. In practice, $M(s)$ is a continuous quantity; to obtain an integer framing, we discretize its variations using the normalization described above, so that a change of approximately one normalized unit in $M(s)$ corresponds to a single unit of framing. We therefore interpret positive increments of $M(s)$ as \emph{positive framing changes}, and negative increments as \emph{negative framing changes}, interpreted over segments of the centroid exceeding a prescribed arc-length scale tied to the reanalysis resolution. In this sense, the topological twist of a band becomes a proxy for the internal reorganization of atmospheric moisture.

A key advantage of this representation is its robustness: smooth deformations of the AR pathway, such as gentle curvature or waviness induced by synoptic flow, primarily affect the geometry of the centroid without necessarily producing a change in $M(s)$ large enough to cross the framing threshold, and hence without altering the integer framing. Only significant internal reorganizations, such as intensification, shear-driven rotation, or substantial weakening, modify $M(s)$ by an amount sufficient to change the framing value. Thus, the framing provides a discrete diagnostic that filters out high-frequency geometric noise and retains the dynamically meaningful evolution of the moisture-transport structure.

The physical interpretation of $M(s)$ is consistent with established diagnostic quantities such as IVT maxima and gradients \cite{GuanWaliser2015}, IWV concentration \cite{Li2024}, tilt and rotation induced by potential-vorticity forcing \cite{Hoskins1985}, and shearing associated with jet streak boundaries \cite{Dacre2019}. Its novelty lies in the fact that it is intrinsically attached to the evolving filament backbone, producing a Lagrangian-like representation of internal changes. When multiple ARs coexist, the framing values along different strands form a vector in $\mathbb{Z}^n$ that evolves over time as the synoptic environment modulates their moisture loading.

In this way, the $\mathbb{Z}^n$ factor in the framed braid group $FB_n$ acquires a physically interpretable role: it records the moisture-driven structural deformation of AR filaments. Together with the geometric interactions captured by the $\sigma_i$ generators, it yields a unified algebraic description that encodes both the pathway geometry and the 
thermodynamic evolution of the moisture corridor.

\subsection{Framed isotopy and atmospheric interpretation}

\begin{figure}[H]
\centering
\includegraphics[width=0.5\linewidth]{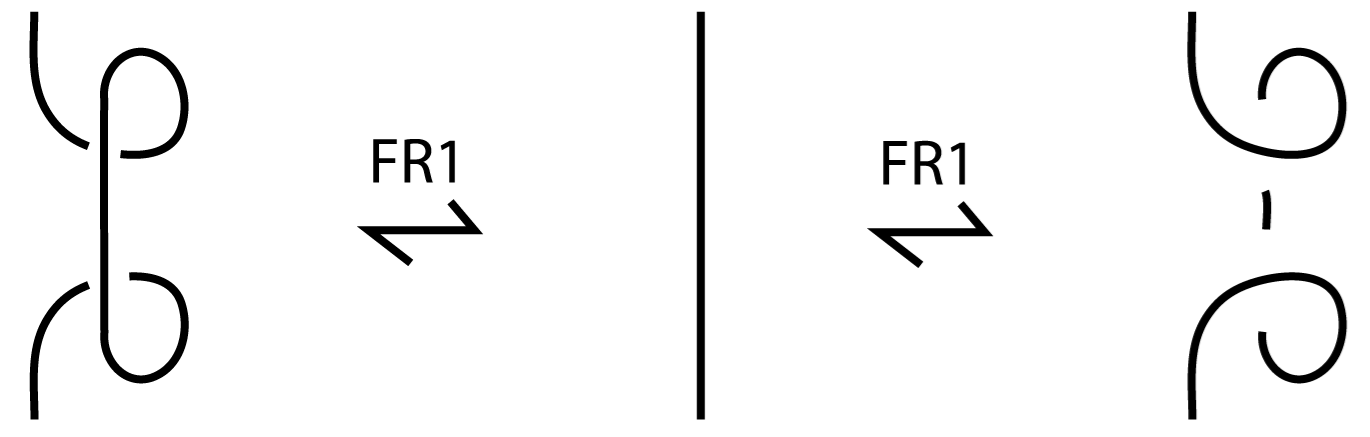}
\caption{The Reidemeister 1 move in the framed setting. The framed type I move (FR1) reflects the correspondence between a physical loop and a unit of framing ($\pm 1$ twist).
}
\label{fig:frR1}
\end{figure}

The topological equivalence of framed braids is governed by a modified set of Reidemeister moves. While moves of type 2 and 3 remain identical to the classical case, the type 1 move is restricted. Specifically, in the framed setting, straightening a curl or a loop is not an ambient isotopy unless we account for the change in the internal twist of the ribbon. As shown in Fig.~\ref{fig:frR1}, the elimination of a loop corresponds to a unit change ($\pm 1$) in the framing integer. 

From a physical perspective, this restriction ensures that the ``internal history'' of an AR filament, such as its sub-grid scale rotation or moisture intensification, is not lost during global track deformations. The framing thus acts as a topological memory of the filament's evolution.

Two framed braids are considered equivalent if one can be smoothly deformed into the other while preserving:
\begin{itemize}
    \item[(i)] the embedded ribbon structure,
    \item[(ii)] the monotonicity in the time direction, and
    \item[(iii)] the disjointness of the bands and their endpoints.
\end{itemize} 

\noindent Such a deformation is called a \emph{framed isotopy}. Informally, the total number of crossings in a diagram may change through local geometric adjustments, but the way strands are connected and the integer framing attached to each strand are preserved.

At the level of diagrams, framed isotopy is generated by moves that realize the relations in $FB_n$:
\begin{enumerate}
    \item small geometric perturbations of the band geometry that do not change the topological type of the projected crossings (ambient isotopy in 
    $\mathbb{R}^2\times[0,1]$),
    \item local rearrangements of independent crossings, corresponding to the Artin braid relations (e.g.\ exchanging the order of two crossings that  involve disjoint pairs of strands), and
    \item redistributions of twist along a band that preserve its total integer framing (for example, sliding a local twist along the strand without creating or removing a full additional twist).
\end{enumerate}

These transformations have clear atmospheric interpretations. Smooth meanderings of an AR centroid induced by the synoptic flow or small geometric adjustments in how nearby filaments weave around each other, correspond to framed isotopies of the underlying braided structure: the centroids deform, but the connectivity between filaments and the discretized 
framing values remain unchanged. By contrast, a coherent intensification or weakening of the moisture corridor or a sustained shear-driven rotation of the cross-sectional structure, modifies the scalar field $M(s)$ in a way that produces an order-one change in the associated integer framing. In our model such events are represented not by isotopies, but by the action of the twist generators $\tau_i$ on the corresponding strand.

Thus, the framing acts as a discrete diagnostic that is invariant under physical deformations which merely advect and gently deform AR filaments, and changes only when the internal moisture organization of a filament is materially reorganized. This separation between framed isotopies (purely geometric evolution at fixed framing) and twist operations (changes in internal structure) will be central in the braid-based description of AR dynamics developed below.

\begin{remark}\rm
We note that braid and framed braid equivalence are only introduced to justify the robustness and physical interpretability of the representation; we do not attempt to classify AR events up to equivalence or construct equivalence-class invariants.    
\end{remark}

\subsection{Framed braidoids: the correct setting for finite AR filaments}
\label{sec:framed_braidoids}

Classical braids assume a fixed number of strands that persist for the entire duration of the time interval. Atmospheric Rivers (ARs), however, do not behave in this idealized way: new filaments may \emph{appear} as coherent moisture plumes, existing filaments may \emph{dissipate} or weaken until they vanish, and single filaments may \emph{split} or \emph{merge} as the synoptic flow reorganizes. Because these processes change the number and connectivity of the filaments, framed braids alone cannot encode the full lifecycle of AR structures.

To model these finite-time and topology-changing events, we adopt the theory of \emph{braidoids}, introduced in \cite{GL1}, as the topological setting. A braidoid is an open braid in $\mathbb{R}^2 \times [0,1]$: its strands may terminate at the boundary planes or end freely in the interior, and may have free endpoints in the interior. Braidoids therefore form the natural topological framework for systems where filaments may appear, disappear, or interact in nonconservative ways. When each strand is further equipped with an integer framing, we obtain a \emph{framed braidoid}, capable of capturing both the geometric evolution and the internal moisture dynamics of AR filaments.

\begin{figure}[H]
\centering
\includegraphics[width=0.13\linewidth]{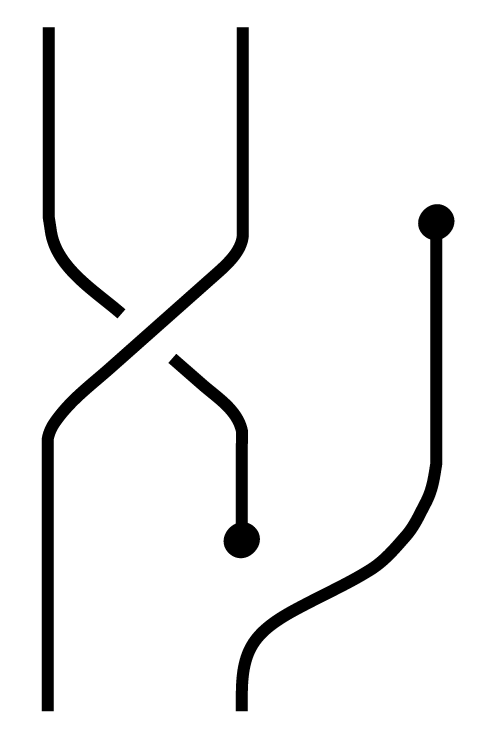}
\caption{A braidoid.}
\label{fig:braidoid_structure}
\end{figure}

\begin{definition}\rm
A \emph{framed braidoid} consists of a braidoid diagram in which each strand is replaced by a band carrying an integer framing. Allowed isotopies must preserve
(i) monotonicity in the time direction, (ii) disjointness of band interiors, (iii) attachment of free endpoints, and (iv) the integer framing on each strand. In particular, any local move that changes the total twist of a strand or that repositions a free endpoint in a way incompatible with braidoid isotopy is forbidden.
\end{definition}

Two families of forbidden moves arise in this setting.  
The first is inherited from classical braidoids: an endpoint may not pass over or under another strand. Physically, this reflects the fact that the head or tail of an AR filament cannot ``jump'' past another filament without an actual dynamical reconnection event.

The second forbidden move is new to the framed setting and is essential for the physical interpretation adopted here: a free endpoint cannot rotate about its band axis in a way that changes its integer framing. Such a rotation would create or remove a full twist, which in the atmospheric model corresponds to a material reorganization of cross-sectional moisture structure. Because AR genesis and dissipation points cannot undergo such rotations without genuine moisture-flux restructuring, these moves are disallowed. This restriction distinguishes framed braidoids from classical framed braids and ensures that framing changes encode physically meaningful variations in $M(s)$ rather than diagrammatic artifacts.

\begin{figure}[H]
\centering
\includegraphics[width=0.45\linewidth]{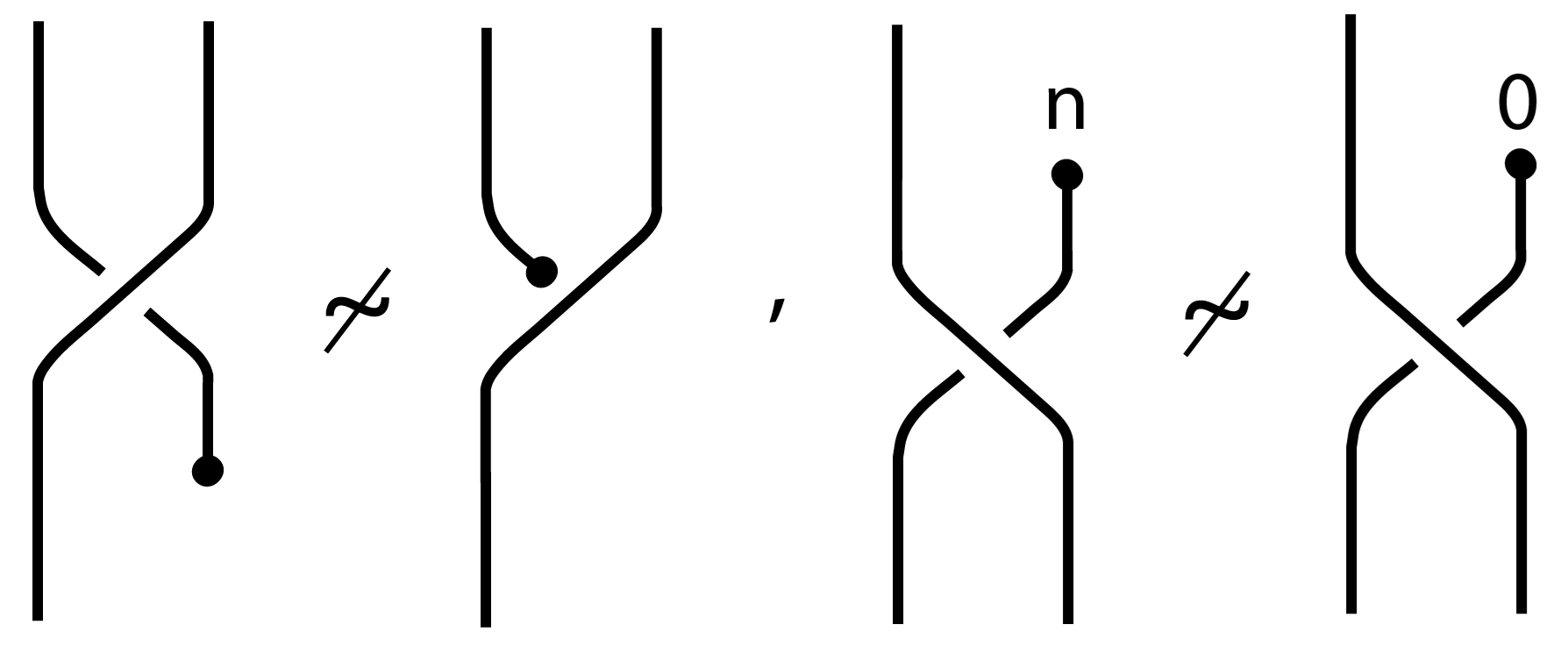}
\caption{Forbidden moves in a framed braidoid.
Left: An endpoint passing under another strand.
Right: Rotational motion of an endpoint changing the integer framing.}
\label{fig:forbidden_move}
\end{figure}

\begin{remark}\rm
The framed braidoid structure introduced here, including the isotopy restriction at endpoints, constitutes a new topological object in the framed setting considered here. A systematic development of framed braidoids, their algebraic presentations and associated invariants will be pursued in a sequel paper.
\end{remark}

\paragraph{Braidoids and knotoids.}
The braidoid formalism is closely related to the theory of \emph{knotoids}, introduced by Turaev \cite{Turaev}. A knotoid is an open knot diagram with two distinguished endpoints, considered up to Reidemeister moves that avoid sliding an endpoint across a crossing. Knotoids provide a diagrammatic counterpart to braidoids: closing a braidoid yields a knotoid diagram, while cutting a knotoid along a monotone height function produces a braidoid. In this sense, braidoids may be viewed as the ``time-parametrized'' version of knotoids. The restrictions on endpoint motion in knotoid theory mirror the forbidden endpoint moves in braidoids and provide additional intuition for the role of free endpoints in representing finite-length AR filaments. The framed extension adopted here simply adds an integer-valued twist structure to this correspondence.

\paragraph{Generators and atmospheric interpretation.}

The elementary algebraic operations that we require for modeling AR evolution are:
\[
\{\sigma_i^{\pm1}\} \cup \{\tau_i\},
\]
where
\begin{itemize}
\item $\sigma_i^{\pm1}$ : strand crossing  
\hfill (lateral reordering of AR centroids),
\item $\tau_i$ : full twist on band $i$  
\hfill (moisture intensification or shear-driven rotation).
\end{itemize}

\begin{remark}\rm 
Although AR genesis, dissipation and branching events may be interpreted diagrammatically as local endpoint creation, endpoint removal, or trivalent splitting/merging, we do not introduce algebraic generators for these moves. Such extensions can be formalized using the theory of knotted trivalent graphs (for example, in Caprau’s work on trivalent graphs \cite{Capr}), but the present model requires only the topological constraints encoded in the braidoid structure 
and the framed endpoint restrictions described above. These are sufficient to model conceptually the finite-time appearance, disappearance, and interaction of AR filaments.
\end{remark}

\subsection{When does a braid suffice?}

Although framed braidoids provide the most general topological representation of AR evolution, framed braids often suffice over moderate time intervals. In
particular, during periods of approximately $12$--$48$ hours, ARs frequently exhibit:
\begin{itemize}
\item a fixed number of coherent filaments,
\item no significant branching or merging, and
\item stable cross-sectional moisture organization.
\end{itemize}
Under these conditions, the topology of the filament network does not change, and the evolution can be fully encoded by a framed braid word in $FB_n$. The only updates required over time correspond to geometric interactions between filaments (recorded by the $\sigma_i$) and internal moisture evolution (encoded in the $\tau_i$).

\bigbreak

In contrast, braidoids become essential once any of the following occur:
\begin{enumerate}
\item \textbf{Genesis or dissipation:} new moisture plumes form or existing filaments terminate;
\item \textbf{Topological reconfiguration:} filaments bifurcate or coalesce in response to the ambient flow;
\item \textbf{Spatial detachment of substructures:} secondary coherent cores or local maxima drift away from the primary filament;
\item \textbf{Extended-time reconstruction:} multi-day AR evolution in which the number, identity, or connectivity of filaments changes.
\end{enumerate}

\section{From AR Track Data to Braids}
\label{sec:braid_extraction}

In this section we describe how pre-identified Atmospheric River (AR) centroid trajectories are converted into oriented braid words that encode their time-ordered geometric interactions. The procedure extracts the sequence of crossings among AR filaments within a finite temporal window, thereby providing a symbolic representation of their mutual entanglement. Regions of elevated braid complexity correspond to intervals in which filaments interact strongly or undergo rapid geometric reconfiguration. Moisture-based framing and branching events are incorporated in Sections~\ref{sec:framed_braids} and \ref{sec:framed_braidoids}.

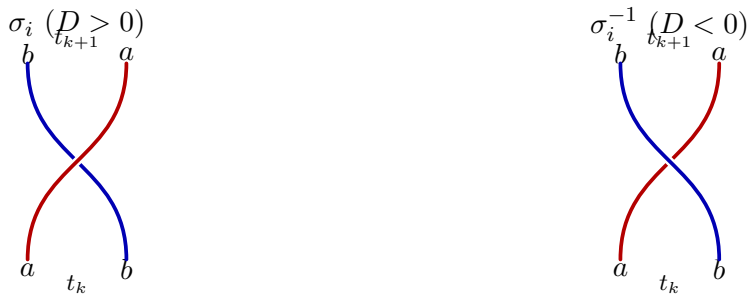
\begin{figure}[H]
\centering
\begin{tikzpicture}[
    scale=1.3,
    strand/.style={line width=1.4pt, line cap=round},
    over/.style={preaction={draw=white, line width=3.2pt}}
]

% --- Left panel: sigma_i (a over b) - POSITIVE CROSSING ---
\begin{scope}[shift={(-3,0)}]
    % labels
    \node at (0,1.4) {$\sigma_i$ ($D>0$)};
    
    % bottom endpoints
    \node at (-0.5,-1.1) {$a$};
    \node at (0.5,-1.1) {$b$};
    % top endpoints
    \node at (-0.5, 1.1) {$b$};
    \node at (0.5, 1.1) {$a$};

    % strand a (blue) goes from R to L (UNDER)
    \draw[strand, blue!70!black]
        (0.5,-1) .. controls (0.5,0) and (-0.5,0) .. (-0.5,1);
    
    % strand b (red) goes from L to R (OVER)
    % This is the strand that starts at the 'a' locus and moves right.
    \draw[strand, red!70!black, over]
        (-0.5,-1) .. controls (-0.5,0) and (0.5,0) .. (0.5,1);
\end{scope}

% --- Right panel: sigma_i^{-1} (b over a) - NEGATIVE CROSSING ---
\begin{scope}[shift={(3,0)}]
    % labels
    \node at (0,1.4) {$\sigma_i^{-1}$ ($D<0$)};
    
    % bottom endpoints
    \node at (-0.5,-1.1) {$a$};
    \node at (0.5,-1.1) {$b$};
    % top endpoints
    \node at (-0.5, 1.1) {$b$};
    \node at (0.5, 1.1) {$a$};

    % strand b (red) goes from L to R (UNDER)
    \draw[strand, red!70!black]
        (-0.5,-1) .. controls (-0.5,0) and (0.5,0) .. (0.5,1);

    % strand a (blue) goes from R to L (OVER)
    % This is the strand that starts at the 'b' locus and moves left.
    \draw[strand, blue!70!black, over]
        (0.5,-1) .. controls (0.5,0) and (-0.5,0) .. (-0.5,1);
\end{scope}

% --- Χρονικές ετικέτες ---
\node[font=\footnotesize] at (-3, 1.25) {$t_{k+1}$};
\node[font=\footnotesize] at (-3, -1.25) {$t_{k}$};
\node[font=\footnotesize] at (3, 1.25) {$t_{k+1}$};
\node[font=\footnotesize] at (3, -1.25) {$t_{k}$};
\end{tikzpicture}
\caption{Detection of strand crossings using the signed oriented area (determinant $D$). A change in longitudinal ordering between times $t_k$ and $t_{k+1}$ produces a crossing. If $D>0$ we assign $\sigma_i$ (left panel); if $D<0$ we assign $\sigma_i^{-1}$ (right panel).}
\label{fig:crossing_orientation}
\end{figure}

The input consists of centroid trajectories 
\[
(x_i(t_k),y_i(t_k)), \qquad i=1,\dots,N,\quad k=0,\dots,T,
\]
sampled at discrete times from an AR catalog or an automated detection algorithm. The key idea is that changes in the longitudinal ordering of centroids signal the occurrence of a crossing, while the \emph{orientation} of the crossing is determined by the sign of a cross-product-like determinant.

\begin{algorithm}[H]
\caption{Oriented braid extraction from AR centroid trajectories}
\DontPrintSemicolon

\KwIn{Centroid trajectories $(x_i(t_k),y_i(t_k))$ for $i=1,\dots,N$, 
\; $k=0,\dots,T$.}

\KwOut{Oriented braid word $\beta$.}

\BlankLine
\textbf{1. Preprocessing and temporal smoothing:}
\Indp
Interpolate or impute missing centroid positions.\;
Apply temporal smoothing (e.g.\ Savitzky--Golay filter) to suppress high-frequency noise.\; Optionally restrict to filaments within a longitudinal window of width $\Delta_{\mathrm{span}}$.\;
\Indm

\BlankLine
\textbf{2. Compute longitudinal ordering:}
\For{$k = 0,\dots,T$}{
    $\mathrm{order}_k \leftarrow \text{argsort}(x_1(t_k),\dots,x_N(t_k))$\;
}

\BlankLine
\textbf{3. Detect crossings between consecutive time steps:}
\For{$k = 0,\dots,T-1$}{
    \For{each adjacent pair $(a,b)$ in $\mathrm{order}_k$}{
        \If{the order of $(a,b)$ is reversed at time $t_{k+1}$}{
            Compute oriented area
            $D = \Delta x_a \Delta y_b - \Delta y_a \Delta x_b$.\;
            \If{$|D| < \varepsilon$}{discard near-degenerate interaction\;}
            \ElseIf{$D > 0$}{append $\sigma_j$ to $\beta$\;}
            \ElseIf{$D < 0$}{append $\sigma_j^{-1}$ to $\beta$\;}
        }
    }
}

\BlankLine
\textbf{4. Return braid word $\beta$.}

\end{algorithm}

Changes in the longitudinal ordering identify potential crossings, but ordering alone does not determine the \emph{orientation}. To assign a consistent signed crossing orientation (the braid generator sign), we compute:
\[
D = \Delta x_a \Delta y_b \;-\; \Delta y_a \Delta x_b,
\]
the signed oriented area (or $2$D cross product) between the displacement vectors of strands $a$ and $b$ over the time interval $[t_k,t_{k+1}]$. Since longitude-latitude coordinates are spherical and expressed in degrees, we compute
$(\Delta x,\Delta y)$ after mapping $(\lambda,\phi)$ to a local Cartesian projection (e.g.\ equirectangular with $\Delta x=\cos(\phi_0)\,\Delta\lambda$, $\Delta y=\Delta\phi$,
or a local tangent-plane projection), where $\phi_0$ is an appropriate reference latitude (e.g.\ the mean latitude of the two strands over $[t_k,t_{k+1}]$). This ensures that the
signed area determinant $D$ is geometrically comparable across latitudes. Accordingly, the tangency threshold $\varepsilon$ is set in the resulting projected units
(and can be converted back to degree-based units for reporting if desired).

If $D>0$, $\sigma_j$ is recorded; if $D<0$, the opposite orientation yields $\sigma_j^{-1}$. Small values of $|D|$ indicate nearly parallel motion or numerically unstable crossings and are discarded using a threshold $\varepsilon$.

The determinant test is both geometrically natural and robust to moderate positional error. It yields a consistent orientation assignment and ensures that the resulting braid word faithfully represents the time-ordered evolution of the AR centroid network.

\subsection{AR track data}

We use a catalogue of Atmospheric River (AR) tracks extracted from the ERA5 reanalysis using the detection and tracking methodology of \cite{GuanWaliser2015, GuanWaliser2019, Rutz2019}. Each six-hourly record corresponds to one AR signature at a given time step and contains:

\begin{itemize}
    \item \verb|time|: timestamp of the detection,
    \item \verb|trackid|: unique identifier for a coherent AR filament,
    \item \verb|centroid_x|, \verb|centroid_y|: longitude and latitude of the IVT-defined centroid,
    \item additional AR properties such as IVT magnitude 
    \cite{Neiman2008, GuanWaliser2015}, detected area, and land interaction.
\end{itemize}

For each \verb|trackid| we obtain a discrete centroid trajectory
\[
t_k \;\longmapsto\; \bigl(x(t_k),\, y(t_k)\bigr),
\qquad k = 0,\dots,T,
\]
providing a time-ordered representation of the pathway of maximum moisture transport over a finite temporal window.

After loading the CSV archive into a \texttt{pandas} DataFrame, we apply the following preprocessing steps: (i) removal of filaments shorter than 12\,hours; 
(ii) forward/backward filling of isolated missing coordinates; (iii) temporal smoothing using a low-order Savitzky--Golay filter to suppress high-frequency positional noise.

To focus on Pacific-origin AR events with strong synoptic coupling to the North American storm track, we restrict attention to the domain
\[
10^\circ\mathrm{N} \le \phi \le 70^\circ\mathrm{N},\qquad
150^\circ\mathrm{E} \le \lambda \le 110^\circ\mathrm{W},
\]
expressed in the $[-180^\circ,180^\circ]$ longitude convention
\cite{RutzSteenburgh2014, Ralph2019, Payne2020}. The resulting regional subset, denoted $df_{\mathrm{reg}}$, forms the basis for all subsequent braid-based diagnostics.

\begin{figure}[H]
\centering
\includegraphics[width=0.7\linewidth]{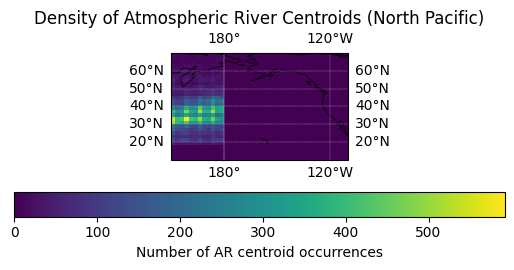}
\caption{Spatial distribution of Atmospheric River centroid positions in the North Pacific regional domain (10°N–70°N, 150°E–110°W). Shading indicates the frequency of centroid occurrence, highlighting dominant AR corridors that form 
the input to braid extraction.}
\label{fig:denistyAR}
\end{figure}

\subsection{Sliding time windows and strand selection}

To analyze local entanglement structure as ARs evolve, we use sliding time windows of fixed duration $T_{\mathrm{win}}$ (typically 48 hours), consistent with established AR lifecycle diagnostics \cite{Dettinger2013, Neiman2008}.
For each central analysis time $t_0$ in a predefined set of timestamps, we define the interval
\[
[t_{\mathrm{start}}, t_{\mathrm{end}}]
=
\Bigl[t_0 - \frac{T_{\mathrm{win}}}{2},\;
      t_0 + \frac{T_{\mathrm{win}}}{2}\Bigr].
\]
Within this window, we extract the subset
\[
df_{\mathrm{win}}
=
\{\text{rows of } df_{\mathrm{reg}} \mid 
  t_{\mathrm{start}} \le \texttt{time} \le t_{\mathrm{end}}\}.
\]

We identify all filaments (distinct \verb|trackid| values) present during the interval and retain at most \verb|max_strands| of them, choosing those with the largest number of sampled time steps.  
This yields a set of
\[
M \;\le\; \texttt{max\_strands}
\]
well-resolved trajectories that are sufficiently complete over the selected window.

Let $\{t_k\}_{k=0}^{T-1}$ denote the sorted timestamps in $df_{\mathrm{win}}$, and let $\{s_i\}_{i=1}^M$ be the retained track IDs. We then construct the matrices
\[
X_{k,i} = x_{s_i}(t_k),
\qquad
Y_{k,i} = y_{s_i}(t_k),
\]
where missing values are filled using forward and backward interpolation (\verb|ffill|/\verb|bfill| in \texttt{pandas}), producing continuous longitude–latitude trajectories for all $M$ strands.

\paragraph{Compact strand clusters.}
Not all $M$ filaments present in the time window exhibit close geometrical or dynamical interaction. To isolate coherent bundles, we compute the mean longitude
\[
\bar{x}_i = \frac{1}{T}\sum_{k=0}^{T-1} X_{k,i},
\]
sort the $M$ strands by $\bar{x}_i$, and search for subsets
\[
S \subseteq \{1,\dots,M\},
\qquad |S| = \texttt{cluster\_size} = N,
\]
whose mean longitudes satisfy the compactness condition
\[
\max_{i \in S} \bar{x}_i - \min_{i \in S} \bar{x}_i
\;\le\; \Delta_{\mathrm{span}},
\]
where $\Delta_{\mathrm{span}}$ is typically chosen around $80^\circ$. This selection step ensures that the chosen strands form a geometrically coherent group consistent with prior studies identifying AR bundles through spatial proximity and shared synoptic forcing \cite{Gorodetskaya2020, Lavers2011}. If no such subset is found for a given window, the window is discarded from further analysis.

In summary:
\begin{itemize}
\item $M$ denotes the number of candidate strands present and sufficiently sampled in the window,
\item \texttt{cluster\_size} $= N$ specifies the number of strands actually used to construct the braid,
\item typically $N < M$, and selection is based on longitude-based compactness.
\end{itemize}

\subsection{Oriented braid extraction from discrete trajectories}

Given the matrices $X,Y$ for a selected cluster of $N$ strands sampled at times $t_0,\dots,t_{T-1}$, we extract an oriented braid word using a discrete ordering-and-crossing procedure. This follows established Lagrangian braid
diagnostics in dynamical systems \cite{Thiffeault2010, FinnThiffeault2006}.

At each time $t_k$, we determine the zonal (longitude) ordering of strands:
\[
\mathrm{order}_k = \operatorname{argsort}\bigl(X_{k,\cdot}\bigr),
\]
which sorts strands from west to east. Comparing $\mathrm{order}_k$ and
$\mathrm{order}_{k+1}$ identifies potential crossings as the system evolves.

\paragraph{Braid axis convention.}
We take time as the braid axis (monotone vertical coordinate). Longitude is used as the ordering coordinate (west–east projection) to determine strand permutations and crossings.

\medskip

For each adjacent pair in $\mathrm{order}_k$, indexed by $(j,j+1)$, define:
\[
(a_{\mathrm{now}}, b_{\mathrm{now}})=
\bigl(\mathrm{order}_k[j],\, \mathrm{order}_k[j{+}1]\bigr),
\qquad
(a_{\mathrm{next}}, b_{\mathrm{next}})=
\bigl(\mathrm{order}_{k+1}[j],\, \mathrm{order}_{k+1}[j{+}1]\bigr).
\]
A swap is detected when
\[
a_{\mathrm{now}} = b_{\mathrm{next}}
\quad\text{and}\quad
b_{\mathrm{now}} = a_{\mathrm{next}},
\]
indicating that strands $a$ and $b$ crossed between $t_k$ and $t_{k+1}$.

\paragraph{Orientation via velocity determinants.}
To assign over/under orientation, we approximate the strand velocities:
\[
\mathbf{v}_a =(\Delta x_a, \Delta y_a)=
\bigl(X_{k+1,a_{\mathrm{now}}}-X_{k,a_{\mathrm{now}}},
      Y_{k+1,a_{\mathrm{now}}}-Y_{k,a_{\mathrm{now}}}\bigr),
\]
\[
\mathbf{v}_b =(\Delta x_b, \Delta y_b)=
\bigl(X_{k+1,b_{\mathrm{now}}}-X_{k,b_{\mathrm{now}}},
      Y_{k+1,b_{\mathrm{now}}}-Y_{k,b_{\mathrm{now}}}\bigr).
\]
We compute the oriented area
\[
D =
\Delta x_a\,\Delta y_b - \Delta y_a\,\Delta x_b,
\]
whose sign distinguishes clockwise from counterclockwise exchange. Small values of $D$ correspond to near-tangential motion and are ignored.

Thus:
\[
D>0 \Rightarrow \sigma_{j},
\qquad
D<0 \Rightarrow \sigma_{j}^{-1}.
\]
(Here $j$ is the braid index in standard one-based notation.)

Scanning all pairs and time steps yields the oriented braid word
\[
\beta =
\sigma_{i_1}^{\varepsilon_1}\sigma_{i_2}^{\varepsilon_2}\cdots\sigma_{i_m}^{\varepsilon_m},
\qquad
\varepsilon_\ell\in\{+1,-1\},
\]
with associated metadata (strand labels, timestamps, determinant values).

\paragraph{Tangency threshold.}
Because centroid positions arise from six-hourly reanalysis fields, small longitude–latitude perturbations may generate spurious crossing signals. We impose a tolerance $\varepsilon$ and accept a crossing only when
$|D|>\varepsilon$. Choosing $\varepsilon$ in the range
$10^{-6}$–$10^{-5}$~deg$^2$ matches typical sub-grid displacement noise. Braid words are stable across this range, indicating that the threshold is both physically and numerically robust.

%%%%%%%%%%%%%%%%%%%%%%%%%%%%%%%%%%%%%%%%%%%%%%%%%%%%%%%%%%%%%
%%%%%%%%%%%%%%%%%%%%%%%%%%%%%%%%%%%%%%%%%%%%%%%%%%%%%%%%%%%%%

\subsection{Braid complexity and selection of representative events}

For each sliding window we extract an oriented braid word $\beta$ describing the geometric interactions of the selected AR filaments. We quantify the associated topological activity using the \emph{crossing count}, defined as the word length
\[
C(t_0) = m,
\]
where $m$ is the total number of generators appearing in $\beta$ (each occurrence of $\sigma_j$ or $\sigma_j^{-1}$ contributes one unit). This measure follows braid–based transport diagnostics \cite{Boyland2000, FinnThiffeault2006}. Larger values of $C(t_0)$ indicate more frequent changes in lateral ordering among AR centroids and therefore enhanced
dynamical interaction.

\begin{remark}\rm
We use word length as a simple interaction proxy; more refined braid-complexity measures (e.g., entropy-based) are possible but beyond scope.    
\end{remark}

To identify the most entangled episodes in a region, we slide the central time $t_0$ over a set of candidate timestamps with a fixed step (typically 10 data steps, e.g.\ $\approx 2.5$~days for six–hourly fields), compute $C(t_0)$ for
each window, and record
\[
C_{\max}=\max_{t_0} C(t_0).
\]
The time $t_0^\ast$ attaining this maximum defines a representative multi–filament interaction event, with associated braid word and strand labels
\[
\bigl(t_0^\ast,\,
\beta^\ast,\,
\{s_i^\ast\},\,
[t_{\mathrm{start}}^\ast,t_{\mathrm{end}}^\ast]\bigr).
\]

\paragraph{Normalization options.}
Because word length depends on the window duration $T_{\mathrm{win}}$ and the sampling interval, it is often useful to consider the normalized quantity
\[
\widehat{C}(t_0) = \frac{C(t_0)}{T_{\mathrm{win}}},
\]
which represents a crossing rate per unit time. In this paper we report $C(t_0)$ in Section~\ref{sec:case_studies}, where we use the complexity scan to identify the maximally entangled three-strand event and analyze this episode, together with a comparison to a larger six-strand AR bundle. The tangency threshold described earlier ensures that these metrics are robust to small positional noise in the trajectory data.

\subsection{Time-longitude braid diagrams and crossing visualization}

To visualize the extracted braid structure, we construct time--longitude diagrams for the selected strands $\{s_i^\ast\}$. Each strand is represented by its discrete longitude sequence
\[
t_k \longmapsto X_{k,i}^\ast,
\qquad k = 0,\dots,T-1,
\]
plotted as a colored curve in the $(t,\lambda)$ plane. This produces a planar geometric braid diagram whose projection is fully consistent with the ordering procedure used in Section~4.

\begin{figure}[H]
\centering
\includegraphics[width=0.6\textwidth]{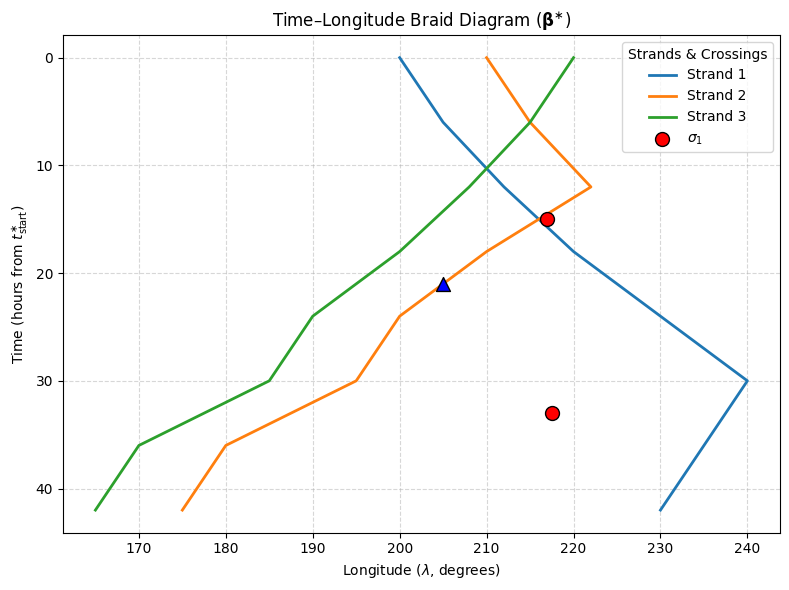}
\caption{
Time–Longitude Braid Diagram for the Representative Event $t_0^\ast$. The figure illustrates the time-ordered evolution of the $N=3$ selected AR strands (curves, $\lambda$ vs. $t$) within the analyzed window $[t_{\mathrm{start}}^\ast, t_{\mathrm{end}}^\ast]$. Markers are superimposed at the approximate crossing locations, with distinct symbols differentiating between the algebraic generators $\sigma_j$ (positive crossings) and $\sigma_j^{-1}$ (negative crossings). This diagram provides the planar geometric projection of the extracted braid word $\beta^\ast$, establishing a direct visual correspondence between the ARs' dynamical interaction and their topological entanglement.
}
\label{fig1:timelon-braid1}
\end{figure}

Using the crossing metadata associated with the braid word $\beta^\ast$, we superimpose markers at the approximate crossing times and longitudes. Distinct symbols denote positive ($\sigma_j$) and negative ($\sigma_j^{-1}$) crossings, providing a direct correspondence between the algebraic word form and the physical exchange of filament positions.

Because all longitudes are expressed in the same $[-180^\circ,180^\circ]$ convention, the plotted curves remain continuous within the regional domain and avoid artificial jumps across the antimeridian. Interactions occurring between discrete sampling times are placed at the midpoint of the interval $[t_k,t_{k+1}]$, which introduces negligible temporal error given the six-hour resolution of the reanalysis fields.

These diagrams offer an intuitive assessment of topological entanglement: tight geometrical clustering and frequent crossings correspond to dynamically active multi-filament interactions. In the next section, we enrich this representation by incorporating moisture-based framing and cross-sectional band structure, enabling a combined geometric and thermodynamic interpretation of AR evolution \cite{Maclennan2025, EirasBarca2023}.

% ==========================================================
\section{From Oriented Braids to Framed Braids for Atmospheric Rivers}
\label{sec:framed_braids}
% ==========================================================

The oriented braid representation of Section~\ref{sec:braid_extraction} encodes changes in lateral positioning among AR centroids, quantifying their geometric entanglement. However, ARs are not one-dimensional filaments: their internal moisture loading and shear strongly influence precipitation outcomes and hydrological impacts. To incorporate this structural information, we enrich each strand of the braid with a \emph{framing} determined by the evolving moisture field of the AR, following the classical theory of framed braids \cite{KoSmo1, Kauffman1991}.

% ----------------------------------------------------------
\subsection{Moisture-based framing}

Let $s_i$ denote the $i$--th selected AR centroid trajectory sampled at discrete times $t_0,\dots,t_{T-1}$. At each time $t_k$ we extract a perpendicular cross–section $\Sigma_i(t_k)$ of the IVT field surrounding $s_i$, computed from the gridded reanalysis IVT$(x,y,t_k)$ using the regional AR mask \cite{GuanWaliser2015, Maclennan2025}. The integrated moisture load is defined as
\begin{equation}
\label{eq:moisture_integral}
M_i(t_k)
=
\int_{\Sigma_i(t_k)} \mathrm{IVT}(x,y,t_k)\,\mathrm{d}A,
\end{equation}
approximated numerically by summing IVT over the detected AR-band pixels in the cross–section. 

\begin{figure}[H]
\centering
\includegraphics[width=0.55\textwidth]{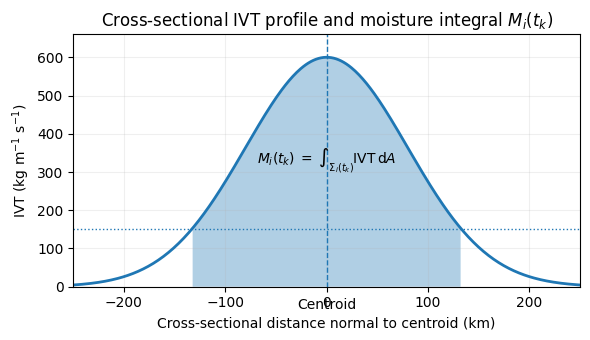}
\caption{
Cross-sectional IVT profile normal to an AR centroid at time $t_k$. 
The shaded area represents the region $\Sigma_i(t_k)$ over which the moisture integral $M_i(t_k)=\int_{\Sigma_i(t_k)} \mathrm{IVT}\, \mathrm{d}A$ is computed. Larger shaded areas correspond to compact, moisture-rich filaments with enhanced cross-sectional gradients, leading to higher framing values in the braided representation.}
\label{fig1:timelon-braid2}
\end{figure}

Large values of $M_i(t_k)$ indicate compact, moisture-rich,
strongly sheared filament segments, whereas reductions reflect weakening, lateral spreading, or partial detrainment of moisture \cite{Neiman2008, EirasBarca2023}.

To obtain a dimensionless and resolution-independent framing sequence, we normalize $M_i(t_k)$ by a characteristic regional scale $\kappa$ (e.g.\ the standard deviation of IVT-area integrals across the window) and set
\[
\tau_i(t_k)
=
\operatorname{round}\!\left(\frac{M_i(t_k)}{\kappa}\right)
\in \mathbb{Z}.
\]
The integer $\tau_i(t_k)$ is interpreted as the framing of strand $i$ at time $t_k$, representing an effective framing of the moist AR band around its centroid in the topological sense, rather than a literal physical rotation. Positive increments $\tau_i(t_{k+1})-\tau_i(t_k)$ reflect moisture accumulation or shear-induced compaction, while negative increments correspond to weakening or transverse broadening.

The sequence $\{\tau_i(t_k)\}_{k=0}^{T-1}$ therefore provides a physically meaningful and topologically robust description of the internal structural evolution of each AR filament.

\begin{remark}\rm
\noindent
We emphasize that only \emph{changes} in the framing, $\Delta\tau_i(k)=\tau_i(t_{k+1})-\tau_i(t_k)$, enter the framed braid representation. Absolute values of $\tau_i(t_k)$ serve as a bookkeeping device, while the topologically meaningful information is carried by their increments, which encode the accumulation or release of internal moisture structure along each filament.
\end{remark}

% ----------------------------------------------------------
\subsection{The framed braid word}
% ----------------------------------------------------------

Let
\[
\beta
=
\sigma_{i_1}^{\varepsilon_1}
\sigma_{i_2}^{\varepsilon_2}
\cdots
\sigma_{i_m}^{\varepsilon_m}
\qquad
(\varepsilon_\ell\in\{\pm1\})
\]
be the oriented braid word extracted from centroid crossings in
Section~\ref{sec:braid_extraction}. Between times $t_k$ and $t_{k+1}$, we define the framing contribution as
\[
\Theta_k
=
\prod_{i=1}^N
\tau_i^{\,\Delta\tau_i(k)},
\qquad
\Delta\tau_i(k)
=
\tau_i(t_{k+1})-\tau_i(t_k),
\]
where $\tau_i$ denotes the twist generator associated with the $i$-th strand of the framed braid group $FB_N$ \cite{KoSmo1}. Positive (resp.\ negative) increments generate positive (resp.\ negative) integer twists.

The full framed braid word representing the AR evolution over the window is then
\[
\beta_{\mathrm{fr}}
=
\Theta_0
\,
\sigma_{i_1}^{\varepsilon_1}
\,
\Theta_1
\,
\sigma_{i_2}^{\varepsilon_2}
\,
\cdots
\,
\Theta_{m-1}
\,
\sigma_{i_m}^{\varepsilon_m}
\,
\Theta_m.
\]

\noindent Here, the generators $\sigma_i^{\pm1}$ encode geometric strand crossings, while the twist generators $\tau_i^{\pm1}$ encode moisture evolution along each filament. The framed braid therefore provides a two-layer representation of AR dynamics, combining positional rearrangement with physically meaningful internal deformation.

\begin{remark}\rm
\noindent
The interleaving of twist elements $\Theta_k$ with geometric generators $\sigma_{i_\ell}^{\varepsilon_\ell}$ reflects the temporal structure of the evolution: framing changes are accumulated continuously in time and are therefore assigned to the intervals between successive centroid crossings. Collecting all twist generators at the beginning or end of the word would obscure this temporal ordering and would no longer correspond to the physical evolution of the AR system.    
\end{remark}

\begin{example}[Framed braid word for a 3-filament AR episode]\rm
Consider a time window in which three AR filaments are present
$(N=3)$ and the geometric braid extraction of Section~\ref{sec:braid_extraction} yields the oriented braid word
\[
\beta = \sigma_1 \,\sigma_2^{-1}.
\]
Thus, between $t_0$ and $t_3$ there are two detected crossings:
a positive crossing of strands $1$ and $2$, followed later by a negative crossing of strands $2$ and $3$.

Suppose the moisture--based framing diagnostic of
Section~\ref{sec:framed_braids} gives the following integer framings for each strand:
\[
\begin{array}{c|cccc}
 & t_0 & t_1 & t_2 & t_3 \\ \hline
\tau_1(t_k) & 0 & 1 & 1 & 2 \\
\tau_2(t_k) & 0 & 0 & 1 & 1 \\
\tau_3(t_k) & 0 & 0 & 0 & -1
\end{array}
\]
Between consecutive sampling times we obtain the framing increments
\[
\Delta \tau_i(k) = \tau_i(t_{k+1}) - \tau_i(t_k), \qquad k=0,1,2,
\]
namely
\[
\begin{array}{c|ccc}
 & k=0 & k=1 & k=2 \\ \hline
\Delta\tau_1(k) & +1 & 0  & +1 \\
\Delta\tau_2(k) & 0  & +1 & 0  \\
\Delta\tau_3(k) & 0  & 0  & -1
\end{array}
\]
The corresponding twist elements in the framed braid group
$FB_3 \cong \mathbb{Z}^3 \rtimes B_3$ are
\[
\Theta_0 = \tau_1^{+1}\tau_2^{0}\tau_3^{0} = \tau_1,\qquad
\Theta_1 = \tau_1^{0}\tau_2^{+1}\tau_3^{0} = \tau_2,\qquad
\Theta_2 = \tau_1^{+1}\tau_2^{0}\tau_3^{-1} = \tau_1\,\tau_3^{-1}.
\]

We assume that the detected crossings occur in distinct intervals and are ordered consistently with the sampling times $t_k$, as enforced by the extraction procedure of Section~\ref{sec:braid_extraction}. The framed braid word that encodes both the geometric and moisture evolution is
\[
\beta_{\mathrm{fr}}
=
\Theta_0\,\sigma_1\,\Theta_1\,\sigma_2^{-1}\,\Theta_2
=
\tau_1\,\sigma_1\,\tau_2\,\sigma_2^{-1}\,\tau_1\,\tau_3^{-1}.
\]

Geometrically, $\sigma_1$ and $\sigma_2^{-1}$ record the exchange of lateral positions among the three centroids. Physically, the twist increments $\tau_1$ and $\tau_2$ between and after these crossings represent episodes of moisture accumulation or shear; induced compaction along the corresponding filaments, while $\tau_3^{-1}$ represents a weakening or broadening of the third filament near the end of the window. Thus $\beta_{\mathrm{fr}}$ compactly encodes, in a single algebraic object, both the sequence of centroid crossings and the internal structural evolution of the AR system.
\end{example}

\subsection{Implementation within the numerical pipeline}

The methodology of Section~\ref{sec:braid_extraction} extends naturally to the framed setting. For each interval $[t_k,t_{k+1}]$ we perform:
\begin{enumerate}
    \item detection of oriented centroid crossings via the determinant criterion,
    \item computation of the cross-sectional moisture load $M_i(t_k)$ and its normalized framing value $\tau_i(t_k)$,
    \item evaluation of framing increments $\Delta\tau_i(k)=\tau_i(t_{k+1})-\tau_i(t_k)$ for each strand,
    \item insertion of the corresponding twist elements
          $\tau_i^{\Delta\tau_i(k)}$ between consecutive geometric crossings.
\end{enumerate}

This yields a time-resolved framed braid word
\[
\beta_{\mathrm{fr}} \in FB_N,
\]
encoding both geometric exchanges of filament positions and the internal moisture evolution of each strand. In
Section~\ref{sec:case_studies} we apply this framework to a highly interactive multi-filament AR event and illustrate how the framed braid structure reveals intensification and weakening signatures that are invisible to centroid geometry
alone.

\subsection{Why Atmospheric Rivers form braidoids rather than braids}
\label{sec:why_braidoids}

As discussed in Section~\ref{sec:framed_braidoids}, ARs do not generally satisfy the structural assumptions of classical braid theory. The number of filaments in a regional domain varies over time: new ARs may form upstream, existing filaments may dissipate after landfall, and coherent corridors may split or merge. These processes correspond to the fundamental braidoid events:
\begin{itemize}
    \item endpoint creation (genesis of a new filament),
    \item endpoint annihilation (dissipation or decay),
    \item branching of a filament into multiple descendants,
    \item merging of nearby filaments into a single moisture corridor.
\end{itemize}

While we may denote these diagrammatic events symbolically (e.g. $\varepsilon^\pm$ for endpoint birth/death and $\eta$ for branching), these symbols serve only as labels of \emph{topological moves}. We do not introduce an
algebraic presentation for braidoids; no such presentation currently exists in the literature. The situation is analogous to knotoid theory, where diagrams with free endpoints are studied topologically without requiring group structures.

Allowing both free endpoints and integer twist leads to a \emph{framed braidoid} description in which the evolution of an AR system may involve symbols representing geometric crossings $\sigma_i^{\pm1}$, moisture-based
twist changes $\tau_j^{\pm1}$, and the appearance, disappearance, or branching of strands. This extended framework naturally reflects the finite lifetime and
structural plasticity of AR filaments.

Although ARs are intrinsically braidoid objects, our sliding–window analysis restricts attention to finite intervals in which a fixed subset of filaments is present throughout. Birth and death events are therefore suppressed, allowing a
well-defined framed braid representation without loss of physical relevance. The full framed braidoid setting, including explicit modelling of genesis, dissipation, splitting and merging, will be developed in future work.

%
%%%%%%%%%%%%%%%%%%%%%%%%%%%%%%%%%%%%%%%%%%%%%%%%%%%%%%%%%%%%%%%%%%%%%%%%%%%%%%%%%%%%%%%%%%%%%%%%%%%%%%%%%%%%%%%%%

\section{Case Studies and Results}\label{sec:case_studies}

In this section we illustrate how braid-based and framed-braid-based diagnostics capture the geometric and moisture-driven interactions of Atmospheric Rivers (ARs) in the North Pacific regional subset. We begin with a sliding-window complexity scan that identifies the most topologically active intervals, followed by a detailed analysis of the three-strand braid with maximal complexity and a comparison with a larger six-strand AR bundle. We then introduce the framed braid representation, which incorporates moisture evolution along each AR filament.

% ---------------------------------------------------------
\subsection{Sliding-Window Complexity Scan}
% ---------------------------------------------------------

We compute the braid complexity $C(t_0)$ for all 48-hour windows using the three-strand procedure described in Section~\ref{sec:braid_extraction}. For each central time $t_0$, the pipeline identifies up to eight candidate strands, selects three geographically co-located filaments, extracts their time–longitude ordering, detects oriented crossings, and computes the length of the corresponding braid word.

\begin{figure}[H]
\centering
\includegraphics[width=\textwidth]{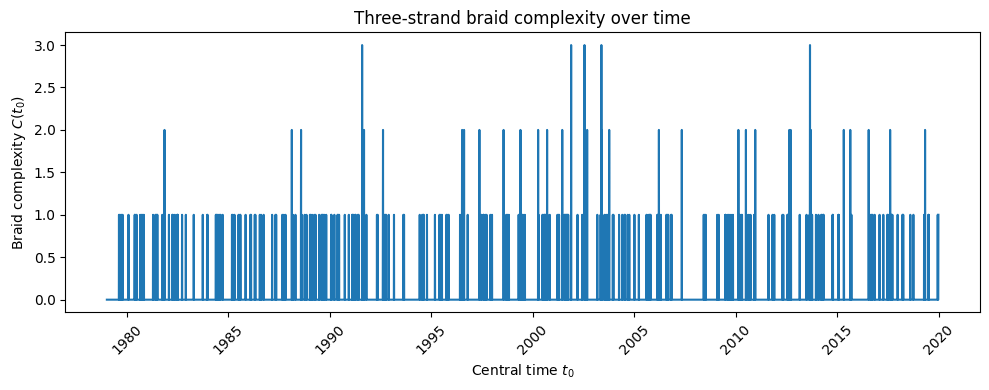}
\caption{
Three-strand braid complexity $C(t_0)$ across all 48-hour windows in the North Pacific regional subset. Most windows satisfy $C(t_0)=0$, while only a small number of windows reach the maximum value $C_{\max}=3$, indicating a uniquely strong geometric entanglement between AR filaments during that interval.
}
\label{fig:complexity-scan-3strand}
\end{figure}

Figure~\ref{fig:complexity-scan-3strand} shows the resulting complexity curve $t_0 \mapsto C(t_0)$. Most windows yield $C(t_0)=0$, indicating no topological interaction. A small number reach $C=1$ or $C=2$, and only a few windows attain the maximum value $C_{\max}=3$. These peak windows identify the most topologically active episodes,  analysed in detail in Section~\ref{sec:3strand-case}.

\begin{figure}[H]
\centering
\includegraphics[width=0.65\textwidth]{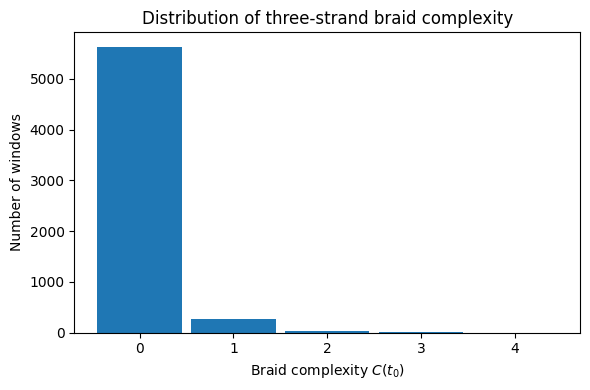}
\caption{
Histogram of three-strand braid complexities $C(t_0)$.  
The distribution is dominated by trivial windows with $C=0$; very few windows yield $C=3$, marking the most topologically active AR interaction in the dataset.
}
\label{fig:complexity-hist}
\end{figure}

The distribution of complexities, shown in 
Figure~\ref{fig:complexity-hist}, is heavily concentrated at $C=0$, confirming that non-trivial braid interactions are relatively rare. Only one window exhibits $C=3$, corresponding to the event examined below.

\begin{figure}[H]
\centering
\includegraphics[width=0.65\textwidth]{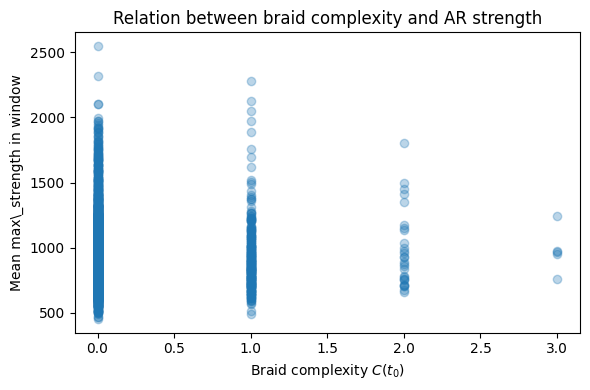}
\caption{
Scatter plot of braid complexity versus mean AR maximum strength within  each 48-hour window. Higher complexities ($C\ge 2$) occur preferentially during periods of  enhanced moisture transport, indicating a connection between dynamical intensity and geometric entanglement.
}
\label{fig:complexity-vs-strength}
\end{figure}

To investigate physical drivers of topological activity, 
Figure~\ref{fig:complexity-vs-strength} compares $C(t_0)$ with the mean maximum strength (IVT magnitude) of ARs within each window. Windows with $C\ge 2$ are observed to occur in periods of enhanced dynamical forcing, suggesting that strong ARs are more likely to exhibit topological entanglement.

% ---------------------------------------------------------
\subsection{Three-Strand Atmospheric River Braid Event}
\label{sec:3strand-case}
% ---------------------------------------------------------

The maximal three-strand braid complexity identified in the North Pacific regional subset occurs in four distinct 48-hour windows, all attaining $C_{\max}=3$ (see Section~\ref{sec:braid_extraction}). In this paper we focus on
the earliest of these events, centered at
\[
t_0^\ast = \text{05 July 1996 15:00 UTC},
\]
which involves the AR tracks
\[
\texttt{39299},\; \texttt{39304},\; \texttt{39312},
\]
and spans the interval
\[
\text{04 July 1996 15:00 UTC} \;\longrightarrow\;
\text{06 July 1996 15:00 UTC}.
\]
During this window the three filaments remain spatially close and undergo three
oriented centroid exchanges, generating the braid word
\[
\beta^\ast = \sigma_2\,\sigma_2^{-1}\,\sigma_2
\]
of length $m=3$.

We emphasize that $C(t_0)$ is the \emph{word length} (crossing count) extracted from the time-ordered ordering changes, rather than a reduced normal form; cancellations such as $\sigma_2\sigma_2^{-1}$ reflect successive back-and-forth exchanges and are physically meaningful in this diagnostic.

\begin{figure}[H]
\centering
\includegraphics[width=0.95\textwidth]{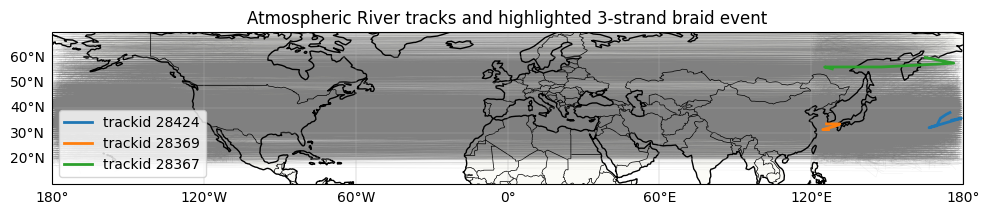}
\caption{
Atmospheric River (AR) trajectories in the North Pacific regional subset during the July 1996 event. Grey curves show all AR centroids in the domain $10^\circ\text{N} \le \phi \le 70^\circ\text{N}$, $120^\circ\text{E} \le \lambda \le 260^\circ\text{E}$. The blue, orange and green curves correspond to trackids, which form a three-strand braid of maximal complexity ($m=3$).
}
\label{fig:map-braid}
\end{figure}

Figure~\ref{fig:map-braid} shows the AR trajectories in the regional domain with the three interacting strands highlighted.

\begin{figure}[H]
\centering
\includegraphics[width=0.8\textwidth]{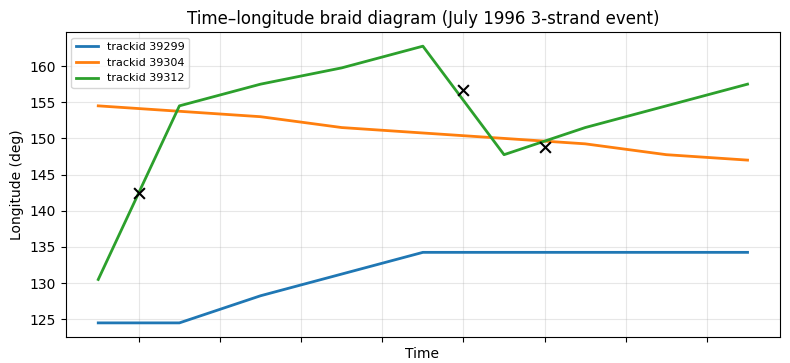}
\caption{
Atmospheric River (AR) trajectories in the North Pacific regional subset during the July 1996 event. Grey curves show all AR centroids in the domain $10^\circ\text{N} \le \phi \le 70^\circ\text{N}$, $120^\circ\text{E} \le \lambda \le 260^\circ\text{E}$. For visualization we use the $[0^\circ,360^\circ]$ longitude convention, which is
equivalent to the $[-180^\circ,180^\circ]$ convention adopted in Section~\ref{sec:braid_extraction} after a standard wrap-around transformation.
}
\label{fig:timelon-braid}
\end{figure}

Figure~\ref{fig:timelon-braid} shows the associated time–longitude braid diagram, with markers indicating the oriented centroid crossings that define the generators of the braid word.

% ---------------------------------------------------------
\subsection{Moisture Evolution and the Framed Braid Structure}
\label{subsec:framed-evolution}

The geometric braid encodes only the lateral rearrangement of AR centroids. To incorporate physically meaningful internal evolution, we construct the framed braid $\beta_{\mathrm{fr}}$ using the moisture-based framing introduced in Section~\ref{sec:framed_braids}. For each strand $s_i$ we compute the normalized moisture integral $\tau_i(t_k)$ via \eqref{eq:moisture_integral}, then record the framing increments
\[
\Delta\tau_i(k)
=
\tau_i(t_{k+1}) - \tau_i(t_k),
\qquad
k = 0,\dots,T-2.
\]
Between two consecutive crossing events, the corresponding twist generators $\tau_i^{\Delta\tau_i(k)}$ are inserted into the framed braid word.

\begin{figure}[H]
\centering
\includegraphics[width=0.85\textwidth]{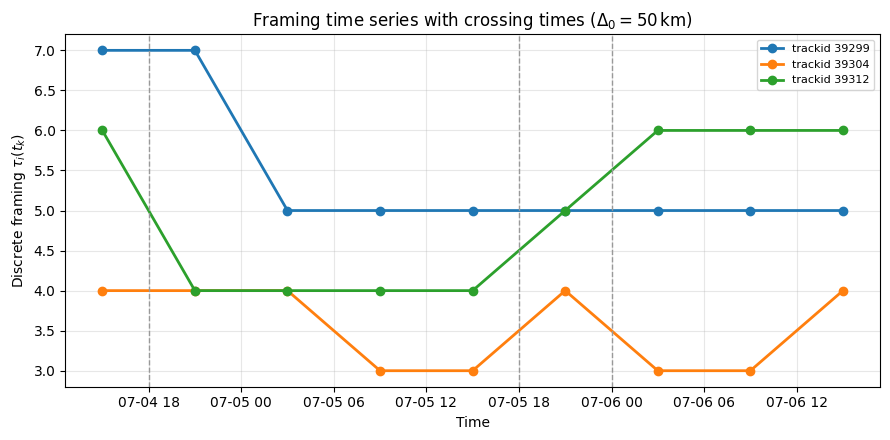}
\caption{
Framing sequences $\tau_i(t)$ for the three interacting filaments in the July~1996 braid event. Vertical dashed lines denote oriented centroid crossings. Moisture intensification in this event occurs near crossing times, suggesting a dynamical coupling between geometric entanglement and internal deformation of AR filaments.
}
\label{fig:framing-timeseries}
\end{figure}

Figure~\ref{fig:framing-timeseries} shows the time evolution of the framing values $\tau_i(t_k)$ for the three strands in the July~1996 event. Vertical dashed lines mark the oriented crossing times. Several moisture-intensification episodes occur near geometric crossings, indicating that dynamical interactions among AR filaments are accompanied by internal moisture reorganization.

% ---------------------------------------------------------
\subsection{Braidoid transitions: appearance, dissipation, and splitting}
% ---------------------------------------------------------

The July 1996 event maintains three coherent filaments throughout the full 48-hour window, allowing a classical framed braid representation. In general, however, AR filament networks do \emph{not} satisfy the assumptions of braid
theory over long times: new filaments may form upstream, existing filaments may decay over land, and coherent corridors may split or merge as the synoptic
environment evolves.

In the braidoid viewpoint (Section~\ref{sec:why_braidoids}), these processes correspond to changes in the number or attachment pattern of strands. While we do not attempt to detect or model such transitions in this paper, it is important to note that AR evolution over multi-day periods is inherently braidoidal.

Our sliding-window framework explicitly avoids birth, dissipation, and splitting by selecting windows in which a fixed subset of filaments persists throughout. This yields a well-defined framed braid representation that captures
the geometric and moisture evolution of the surviving strands. A full treatment of appearance, decay, and branching phenomena will be developed in a sequel paper, where the framed braidoid structure is the natural topological object.

% ---------------------------------------------------------
\subsection{Toward topological early‐warning metrics}
% ---------------------------------------------------------

The observed alignment between geometric crossings and moisture intensification indicates that \emph{topological instability} captures aspects of AR reconfiguration that precede major structural changes in the filament network. In particular, framed braid diagnostics highlight time intervals in which moisture accumulation and geometric entanglement co‐occur, suggesting that braid‐based measures may act as early indicators of dynamically driven precipitation risk.

\paragraph*{Quantitative topological indicators.}
The braid and framed-braid structures naturally give rise to several diagnostic quantities:

\begin{itemize}
\item \emph{Oriented braid complexity} 
      $C_{\mathrm{orb}} = \mathrm{length}(\beta)$,
      measuring geometric reordering via centroid crossings.

\item \emph{Framing variation}
      $V_{\mathrm{frame}} = \sum_{i,k} |\Delta\tau_i(k)|$,
      capturing moisture-driven deformation along the strands.

\item \emph{Moisture Twist Index (MTI)}
      $ \mathrm{MTI}= V_{\mathrm{frame}}/(T-1) $,
      a normalized estimate of internal moisture evolution.

\item \emph{Coupled geometric–moisture complexity}
      $C_{\mathrm{coupled}} = C_{\mathrm{orb}} + V_{\mathrm{frame}}$,
      highlighting windows where geometric interaction and moisture
      intensification co-occur.
\end{itemize}

These indicators are not developed as predictive tools in this paper, but they illustrate how topological structure can be quantified and motivate the early-warning perspective discussed above.

These results motivate the development of topological early‐warning metrics that combine braid complexity, framing variability, and crossing timing. Such metrics would quantify abrupt reorganizations in both geometry and moisture transport, providing complementary information to traditional Eulerian predictors. Extensions of this type, including probabilistic and multi‐scale formulations, will be pursued in future work.

\subsection{Framing Indicators for the July 1996 Event}
\label{subsec:framing_indicators_july1996}

To assess the diagnostic and predictive potential of the topological framework, we evaluate the framing-based indicators introduced earlier—namely, the \emph{Framing Variation} ($V_{\mathrm{frame}}$), the \emph{Moisture Twist Index} (MTI), and the \emph{Coupled Complexity} ($C_{\mathrm{coupled}}$)—for the three-strand Atmospheric River event of 5--7~July~1996.

The \emph{Framing Variation} quantifies the total moisture-driven deformation along the AR filaments:
\[
V_{\mathrm{frame}} = \sum_{i,k} |\Delta\tau_i(k)|,
\]
while the \emph{Moisture Twist Index (MTI)} provides a normalized version:
\[
\mathrm{MTI} = \frac{V_{\mathrm{frame}}}{T - 1},
\]
where $T$ is the number of time steps. To jointly capture geometric reorganization and internal moisture evolution, we compute the \emph{Coupled Complexity}:
\[
C_{\mathrm{coupled}} = C_{\mathrm{orb}} + V_{\mathrm{frame}}.
\]

For the July~1996 case, we find $C_{\mathrm{orb}} = 3$, indicating three strictly oriented centroid crossings. The framing variation reaches $V_{\mathrm{frame}} = 19$, with a resulting $\mathrm{MTI} = 2.375$. The coupled index is therefore $C_{\mathrm{coupled}} = 22$, revealing that internal moisture reorganization dominates the topological activity during this event. The ratio $V_{\mathrm{frame}} / C_{\mathrm{orb}} \approx 6.3$ emphasizes that moisture intensification outweighs geometric entanglement.

Time-resolved plots of $\tau_i(t)$ show that framing changes frequently occur within a few hours of strand crossings, suggesting that geometric interaction and moisture reorganization are dynamically coupled. These co-occurring signals support the interpretation of $V_{\mathrm{frame}}$ and $C_{\mathrm{coupled}}$ as potential \emph{early indicators} of AR intensification or structural transition.

This analysis illustrates that topological indicators can highlight physically meaningful precursors to AR evolution, motivating further development of braid-based diagnostics for operational and climatological applications.

% ---------------------------------------------------------
\subsection{Extensions: pseudo crossings and trivalent interactions}
% ---------------------------------------------------------

\paragraph*{Pseudo crossings and uncertainty.}
Throughout this study, all detected centroid exchanges are treated as classical crossings with a determined over/under orientation. However, coarse six‐hourly sampling and sub‐grid variability may generate apparent order reversals for which the orientation cannot be resolved reliably. Such ambiguous events are naturally modelled using \emph{pseudo crossings}, giving rise to pseudo braids and \emph{pseudo framed braids}. For more information on pseudo knots the reader is referred to \cite{Diamantis1, Diamantis2}. This extension introduces a principled way to represent observational uncertainty within the topological framework and connects directly to recent advances in probabilistic pseudo knot theory.

\paragraph*{Trivalent interactions and filament reconnection.}
Beyond simple appearance and dissipation, ARs frequently undergo mergers and splits that form genuine Y‐shaped junctions. These events cannot be encoded within the classical framed braid group and are only partially accommodated within the framed braidoid setting. Incorporating trivalent nodes yields a topological \emph{braid–graph} structure, where edges carry braid or framed braid data and vertices represent branching or merging. Such hybrid topological objects offer a mathematically precise foundation for future
descriptions of AR filament reconnection and for quantifying the structural reorganization of moisture pathways.

%
%%%%%%%%%%%%%%%%%%%%%%%%%%%%%%%%%%%%%%%%%%%%%%%%%%%%%%%%%%%%%%%%%%%%%%%%%%%%%%%%%%%%%%%%%%%%%%%%%%%%%%%%%%%%%%%%%

\section{Topological Interpretation and Implications}

This work introduced a topological framework for analyzing Atmospheric Rivers (ARs) using oriented braids, framed braids, and framed braidoids. By treating AR filaments as moisture-carrying strands whose centroids and internal structure evolve in time, the framework unifies geometric interaction and physical intensity into a single algebraic object. This perspective complements dynamical and statistical approaches and offers a compact language for describing structural features of AR evolution.

% --------------------------------------------------------------------
\subsection{Conceptual significance of the braid representation}
% --------------------------------------------------------------------

A key insight of the braid formulation is that ARs behave not as isolated features but as \emph{interacting filaments}.  
Their trajectories may cross, twist, merge, or fragment, and each of these events carries meteorological meaning.  
Crossings often arise from differential advection and jet reorganization, while framing changes reflect internal moisture redistribution and shear.

Encoding these interactions as algebraic generators produces a concise yet expressive representation of AR evolution.  
Unlike classical Eulerian or Lagrangian diagnostics, braid-based methods capture the \emph{relative arrangement} of filaments and the \emph{temporal ordering} of their interactions. This provides a structural viewpoint that reveals reorganizations in the AR network that may not be apparent from traditional scalar or trajectory-based
measures.

% --------------------------------------------------------------------
\subsection{Predictive potential for extreme rainfall}
% --------------------------------------------------------------------

Our case study suggests that several braid-derived quantities may exhibit \emph{precursory or concurrent signatures} associated with high-impact precipitation events. In particular:
\begin{itemize}
    \item spikes in oriented braid complexity coincide with sharp curvature changes and synoptic reorganization,
    \item positive framing jumps reflect rapid IVT intensification and moisture loading,
    \item braidoid events such as birth, death, or splitting are conceptually associated with structural fragmentation,
    \item the coupled topological complexity index identifies intervals where geometric and moisture-driven reconfiguration co-occur.
\end{itemize}

These results suggest that braid-based diagnostics have the potential to complement standard predictors such as IVT magnitude, AR width, or centroid curvature, and may enhance early-warning methodologies when combined with ensemble forecasts or machine-learning systems. We emphasize that these observations are exploratory and hypothesis-generating, and do not yet constitute a validated predictive framework.

% --------------------------------------------------------------------
\subsection{Relationship to dynamical frameworks}
% --------------------------------------------------------------------

The topological viewpoint does not replace existing dynamical or statistical frameworks; rather, it provides a complementary structural description. Lagrangian Coherent Structure methods identify attracting and repelling
manifolds in the flow, whereas braid-based representations capture the \emph{interaction topology} between coherent filaments as they evolve and exchange relative position over time. Potential vorticity fields describe the large-scale steering environment in which Atmospheric Rivers are embedded,
while variations in framing encode internal moisture evolution and structural reorganization along individual AR pathways. Similarly, persistent homology extracts topological signatures from scalar fields at fixed times, whereas
braid theory focuses on the time-ordered interactions of moving trajectories. In this sense, the braid framework occupies a distinct structural niche: it is designed to isolate interaction-driven reorganization, complementing both dynamical diagnostics and scalar-field–based topological tools.

% --------------------------------------------------------------------

\subsection{Limitations}
% --------------------------------------------------------------------

Several limitations of the present framework point naturally toward further refinement. The extraction of Atmospheric River tracks depends on IVT thresholds and segmentation algorithms, and alternative tracking methodologies may yield
centroid trajectories with different stability properties, potentially affecting the resulting braid structure. The estimation of framing relies on cross-sectional proxies for moisture content; more physically grounded definitions, such as isentropic or flux-aligned IVT integrals, may improve the
interpretability and accuracy of the framing variable. In addition, braidoid events corresponding to filament birth, dissipation, or splitting are currently identified using heuristic criteria; automated filament tracking or
machine-learning–based segmentation could increase robustness and consistency. Finally, braid entropy and related long-window topological indicators were not computed here, as they typically require long-lived trajectories and therefore
have limited applicability to short or rapidly evolving AR episodes.

Despite these limitations, our exploratory analysis suggests that the primary topological features identified by the braid and framed-braid representations are robust under reasonable methodological variation.

% --------------------------------------------------------------------
\subsection{Future directions}
% --------------------------------------------------------------------

The framework introduced here opens several promising directions for future research. One natural extension is the integration of braid and framing indicators into numerical weather prediction workflows, where their real-time
extraction from ensemble forecasts could contribute to early-warning systems for AR-driven flood risk. From a methodological perspective, combining braid representations with multi-scale topological tools such as persistent homology
or scale-space topology may yield complementary views of moisture transport structure across spatial and temporal resolutions. The inherent uncertainty in AR detection and trajectory estimation also motivates the development of stochastic or probabilistic braid models, providing a principled way to represent ensemble spread and ambiguous filament interactions. In parallel, topological indicators derived from the braid framework, including $C_{\mathrm{orb}}$ and $V_{\mathrm{frame}}$, may serve as physically interpretable features for machine-learning approaches to prediction, clustering, or regime classification. Finally, extending the analysis to climatological timescales opens the possibility of using long-term statistics of braid entropy or framing variability to detect teleconnections, regime shifts, or climate-driven reorganizations in Atmospheric River behavior.

\section{Conclusion}

We have introduced a topologically grounded framework for representing Atmospheric Rivers using oriented braids, framed braids, and framed braidoids. By treating ARs as moisture-bearing filaments whose centroids and internal
structure evolve in time, the framework encodes both geometric interaction and moisture-driven deformation within a unified algebraic representation.

Through case studies, we have shown that braid-based indicators, including oriented braid complexity and framing variation, capture episodes of structural reorganization that are not readily apparent from centroid geometry or scalar intensity measures alone. These results suggest that braid-based topology provides a complementary structural diagnostic that enriches existing dynamical and statistical approaches to Atmospheric River analysis.

More broadly, the proposed methodology offers a general topological framework for studying interacting filamentary structures in geophysical flows. By focusing on interaction-driven reorganization and internal structural evolution, braid-based representations may provide new insight into the
dynamics of coherent structures in complex, time-dependent systems.

\end{document}